\documentclass[12pt]{article}

\usepackage{graphics}
\usepackage{latexsym}

\newcommand{\bartq}[1]{}

\usepackage{amsmath}

\newcommand{\ab}{\allowbreak}
\newcommand{\klk}{, \ldots ,}

\newcommand{\fobetween}{\mbox{${\sf FO}({\sf between})$}}

\newcommand{\foc}{\mbox{${\sf FO}(+,\ab \times,\ab =,\ab 0,\ab 1)$}}
\newcommand{\fop}{\mbox{${\sf FO}(+,\ab \times,\ab =,\ab <,\ab 0,\ab 1)$}}

\newcommand{\foeq}{\mbox{${\sf FO}(+,\ab \times,\ab 0,\ab 1,\ab =)$}}

\newcommand{\ignore}[1]{}

\newcommand{\fopr}[1]{\mbox{${\sf FO}(+,\ab \times,\ab =,\ab <,\ab 0,\ab 1,\ab #1)$}}

\newcommand{\focr}[1]{\mbox{${\sf FO}(+,\ab \times,\ab =,\ab 0,\ab 1,\ab #1)$}}

\newcommand{\mathfrak}[1]{{?? #1}}

\def\squareforqed{\hbox{\rlap{$\sqcap$}$\sqcup$}}
\def\qed{\ifmmode\squareforqed\else{\unskip\nobreak\hfil
\penalty50\hskip1em\null\nobreak\hfil\squareforqed
\parfillskip=0pt\finalhyphendemerits=0\endgraf}\fi}

\newcommand{\R}{{\rm {\bf R}}}
\newcommand{\C}{{\rm {\bf C}}}
\newcommand{\Q}{{\rm {\bf Q}}}

\newcommand{\Zpos}{{\rm {\bf Z}}^+}

\newcommand{\A}{{\rm {\bf A}}}
\newcommand{\N}{{\rm {\bf N}}}
\newcommand{\Pe}{{\rm {\bf P}}}

\newtheorem{theorem}{Theorem}

\newtheorem{definition}{Definition}

\begin{document}

\bibliographystyle{plain}

\sloppy

\title{The evaluation of geometric queries:\\ constraint databases and quantifier elimination}
\author{\begin{tabular}{rl}
\emph{Marc Giusti}, &\'Ecole Polytechnique, France \\ \emph{Joos
Heintz}, & University of Buenos Aires, Argentina\\ \emph{Bart
Kuijpers}, & Hasselt University, Belgium\end{tabular}}

\date{}
\maketitle

\begin{abstract}
We model the algorithmic task of geometric elimination (e.g.,
quantifier elimination in the elementary field theories of real
and complex numbers) by means of certain constraint database
queries, called \emph{geometric queries}. As a particular case of
such a geometric elimination task, we consider \emph{sample point
queries}. We show exponential lower complexity bounds for
evaluating geometric queries in the general and in the particular
case of sample point queries. Although this paper is of
theoretical nature, its aim is to explore the possibilities and
(complexity-)limits of computer implemented query evaluation
algorithms for \emph{Constraint Databases}, based on the
principles of the most advanced geometric elimination procedures
and their implementations, like, e.g., the software package
"Kronecker" (see \cite{gls}). \emph{This paper is based on
\cite{heku} and is only a draft specially prepared for CESSI 2006,
representing work in progress of the authors. It is not aimed for
publication in the present form.}
\end{abstract}

%

\section{Introduction and summary}

The framework of \emph{constraint databases} was introduced in
1990 by Kanellakis, Kuper and Revesz~\cite{kkr95} as a
generalization of the relational database model. Here, a database
consists of a finite number of generalized relations, rather than
classical relations. When we consider constraint databases over
the real numbers, a generalized relation is finitely represented
by a Boolean combination of polynomial equalities and inequalities
over the reals. These so called \emph{constraint formulas}
finitely represent possibly infinite sets in some real space
$\R^n$. Therefore, the constraint database model provides an
elegant and powerful model for applications that deal with
infinite sets of points in some finite dimensional real space, and
is hence well-suited for modelling, e.g., \emph{spatial
databases}. For example, the spatial relation consisting of the
set of points on the northern hemisphere together with the points
on the equator of the unit sphere in the three-dimensional space
$\R^3$ can be represented by the constraint formula $x^2+y^2+z^2
=1 \mathrel{\land} z\geq 0$.

The constraint model has been extensively studied by now and
various logic-based query languages have been
considered~\cite{cdbook}. First-order logic over the reals, \fop,
augmented with relation names to address (generalized) relations
in the input database, is the standard query language for
constraint databases. As an example we may consider, for an input
database, which contains just one ternary relation represented by
$S$, the query expressed by the  \fopr{S}-sentence $$(\exists
r)(\forall x)(\forall y)(\forall z)(S(x,y,z)\rightarrow
x^2+y^2+z^2<r^2).$$ This query expresses that the
three-dimensional spatial relation $S$ is bounded.

The standard way to evaluate this query on a particular database,
e.g.,  the hemisphere above, consists in replacing the
subexpression $S(x,y,z)$
in the query expression by the formula $x^2+y^2+z^2 =1
\mathrel{\land} z\geq 0$ and next to eliminate from the resulting
\fop-formula the quantifiers that were introduced by the query
expression. In our example, this adds up to eliminating the
quantifiers from $$(\exists r)(\forall x)(\forall y)(\forall
z)((x^2+y^2+z^2 =1 \mathrel{\land} z\geq 0)\rightarrow
x^2+y^2+z^2<r^2),$$ which would result in the value \emph{true}.

In this paper, we do not consider the traditional application
domains of constraint databases, such as spatial databases, but
focus on a completely different domain, namely \emph{geometric
elimination theory}.
Hereto, we extend the constraint database model in the sense that
we allow databases also to contain functions, rather than only
relations. In this paper, a typical  (input) database schema will
be of the form $$(R_1,...,R_r;F_1,...,F_s)$$ with relation names
$R_i$ ($i=1,...,r$) and function names $F_j$ ($j=1,...,s$). The
relation names are interpreted, following the context, by
algebraic or semi-algebraic sets and the function names by
polynomial, or exceptionally, rational  functions defined over the
complex or real numbers.

The reason to include function symbols is two-fold. Firstly,
functions appear naturally as byproducts of quantifier-elimination
procedures and it is therefore suitable to consider them as in- or
outputs of such algorithms. So, it is natural to include them if
we want to model appropriately the new application domain of
geometric elimination theory. Typical examples of such functions
are the determinant and the resultants of systems of $n$
homogeneous equations in $n$ unknowns in the linear and non-linear
case, respectively.

The second reason is based on a  complexity argument and explains
why it does not suffice to represent a $k$-ary function just by a
$(k+1)$-ary relation that stores the graph of the function. When
we extend the constraint database model with functions and
likewise extend the first-order query language including function
symbols for the representation of  input functions, we are
sometimes able to write queries more economically with respect to
the number of quantifiers. This leads in turn to more efficient
evaluation of these queries. As an example consider a input schema
containing a unary function symbol $F$. The first-order query
formula

$$y=F(F(x)) $$ of \fopr{F}\ defines, for each interpretation of
the symbol $F$ by a unary real-valued function $f$,
 all tuples $(x,y)$
of real numbers satisfying $y=f^2(x)$. On the other hand, if we
model the function symbol $F$ by means of the graph of $f$, i.e.,
using a database schema  containing a binary relation symbol $R$,
then the relation $y=f^2(x)$ becomes first-order expressible  by
the \fopr{R}-query $$(\exists z)(R(x,z)\land R(z,y)),$$ which
contains a quantifier. Observe that the evaluation of this query
cannot be done directly, it requires the elimination of this
quantifier.

First-order logic over the real or complex numbers extended with
relation and function names, used to address input relations and
functions, allows to define output relations in the traditional
way. On the other hand, the creation of output functions requires
an extension of first-order logic by special terms in order to
\emph{specify}  these functions. To illustrate this, we consider
the following example. Let be given the
\fopr{F_{11},...,F_{nn},F}-formula $$\displaylines{\quad (\exists
x_1)\cdots(\exists
x_n)(\bigwedge_{i=1}^{n}F_{i1}(u_{1},...,u_{m})x_1+\cdots +
F_{in}(u_{1},...,u_{n})x_n =0 \hfill{}(\dagger)\cr\hfill{}\land
\bigvee_{i=1}^n x_i\not=0 )\leftrightarrow
F(u_{1},...,u_{m})=0,\quad \cr}$$
in which the the input schema is given by the function names
$F_{ij}$ ($i,j=1,...,n$), representing polynomial input functions
$f_{ij}(u_1,...,u_m)$ and where the output schema is given by the
term $F$ representing a polynomial  function $f(u_1,...,u_m)$. The
formula $(\dagger)$ may be interpreted as a specification of the
function $f(u_1,...,u_m)$ by the requirement that the condition
 $F(u_{1},...,u_{m})=0$  in $(\dagger)$ reflects that
the linear homogeneous equation system given by the matrix
$\Phi(u_{1},...,u_{m})=(f_{ij}(u_{1},...,u_{m}))_{1\leq i,j\leq
n}$ has a non--zero solution. An example function that satisfies
this specification is the determinant of the matrix $\Phi$.
Another example of a function that satisfies  the specification
$(\dagger)$ is the square of the determinant. In Section
\ref{ssec:qe-as-geometric}, we will discuss generalizations of
this example to non-linear systems of homogeneous equations and it
turns out that possible interpretations of the output functions
are resultants.

We remark that the variables $u_{1},...,u_{m}$ and $x_1,...,x_n$
play a different r\^ole in the expression
 $(\dagger)$. Therefore, we shall use the following terminology:
 we call $u_{1},...,u_{m}$ \emph{parameters} and
 $x_1,...,x_n$ \emph{variables}.
The idea behind this distinction is the following. Suppose we are
given a concrete database instance $(f_{11}(u_1,...,u_m), ...,
f_{nn}(u_1,...,u_m))$ of the the database schema
$(F_{11}(u_1,...,u_m), ..., F_{nn}(u_1,...,u_m))$.\footnote{Or
better?: Suppose we are given a concrete algebraic family of
database instances $(f_{11}(u_1,...,u_m), ...,
f_{nn}(u_1,...,u_m))$, parameterized by $u_1,...,u_m$, over the
the database schema $(F_{11}, ..., F_{nn})$. This way we avoid
using parameterized schemas, which are little bit strange. See
also next page.} This instance gives rise to a new database
containing a single $(m+n)$-ary relation described by the
expression
$$\bigwedge_{i=1}^n\sum_{j=1}^m f_{ij}(u_1,...,u_m)x_j=0.$$ This
relation can be viewed as an algebraic family, which is
parameterized by $u_1,...,u_m$, of database instances consisting
of a single $n$-ary relation in the variables $x_1,...,x_n$.

 The expression $(\dagger)$
is exemplary for the way we shall specify  output functions in
this paper. Thus, output functions will typically take parameters
as arguments.

We also remark that specifications like  $(\dagger)$ are no longer
first-order formulas, but rather higher-order expressions. On the
other hand, we may view these expressions as queries in our new
constraint database model (extended with function symbols).
Indeed,  a (constraint database) \emph{query} is usually defined
as a partial computable function that transforms a given *
\bartq{* = family of?} constraint database instance (over a
previously fixed input schema) into a new constraint database
instance (over an also previously fixed output schema). Observe
that the example described by the expression $(\dagger)$ fits in
this setup.  Indeed, for a given instance of the input symbols
$F_{ij}$ ($i,j=1,...,n$) any instantiation of the function symbol
$F$, satisfying $(\dagger)$,
 can be considered as an output of the corresponding query.

In the sequel, any algorithm that produces an output satisfying a
specification  like $(\dagger)$ will be considered as a query
evaluation algorithm.

Once the constraint database model is extended with functions *,
\bartq{*= and parametrization} the constraint database formalism
may  be used to describe the algorithmic task of geometric
elimination. Hereto, we introduce the notion of \emph{geometric
query}, which is based on the distinction between parameters and
variables, considered before. The idea is to apply queries, which
are specified as above,  to algebraically parameterized families
of input databases. Roughly speaking, a geometric query is a
transformation of algebraic families of constraint databases that
is parameterization independent.
 In this sense, the query which transforms an given input
instance $(f_{11}(u_1,...,u_m), ..., f_{nn}(u_1,...,u_m))$ of the
the previous database schema $(F_{11}(u_1,...,u_m), ...,
F_{nn}(u_1,...,u_m))$ into the determinant of the matrix
$\Phi=(f_{ij}(u_1,...,u_m))_{1\leq i,j\leq n}$, satisfies the
specification $(\dagger)$ and  is geometric.

A central  contribution of this paper is the conclusion  that in
the largest possible sense, any known or thinkable geometric
elimination procedure can be modeled  as the evaluation of a
suitable geometric query. This allows us now to use
\emph{descriptive} specifications in order to formulate basic
tasks of geometric elimination theory. Note that the descriptive
specifications are much more flexible and problem adaptive than
the traditional operative ones as, e.g., the task of the
computation of a certain determinant, resultant or Chow form. In
order to make the statement of this conclusion precise, we have to
concretize  the data structures which  will be used  to represent
the databases which occur as inputs and outputs of geometric query
evaluation algorithms, i.e., of elimination procedures.

In view of progress made in \cite{} concerning the complexity of
geometric elimination algorithms, we shall describe the relations
and functions of a given database instance by the complexity model
of  essentially division-free boolean-arithmetic and arithmetic
circuits, respectively (see Section \ref{} and \cite{} for details
on this model). By the way, we observe that the classical
representation of database instances by quantifier-free
\fop-formulas (or \foc-formulas) is contained in this complexity
model.

With this conceptual tool at hand, we shall be able  to certify
the intrinsic non-polynomial character of geometric elimination
(Theorem \ref{}). However, this complexity result does not exclude
that particular elimination tasks, as, e.g., the evaluation of
certain resultants, would be solvable  in polynomial time.
Motivated by practical applications, we shall therefore consider
the particular elimination task of evaluating \emph{sample point
queries}.

The classical data model for constraint databases \cite{cdbook}
does not support data exploration and local visualization. Indeed,
a quantifier-free formula in disjunctive normal form, describing
the output of a query, allows the answering of, for instance, the
membership question, but it is does not allow an
easy\footnote{Remark: By \emph{definable choice}\cite{vddries} you
can produce sample points in FO. The correctness of this satement
is in the easy.} exhibition of the output, by, e.g., the
production of \emph{sample points}, or, for low dimensions, a
visualization of the output. To increase the tangibility of the
output, we suggest considering a new type of query that produces
sample points. Furthermore, it could be desirable to support an
exploration of the neighborhood of a given sample point. Both
aspects, namely finding a sample point and exploration of its
neighborhood, encounter a simple expression in case of rationally
parameterizable algebraic or semi-algebraic varieties. Therefore,
we shall extend the concept of sample point query to queries that
return rationally parameterized families of polynomial functions
as output. Such queries will be called \emph{extended sample point
queries}. In Section \ref{sec:lb-samplepoint}, we shall prove that
extended sample point queries, associated to first-order formulas
containing a \emph{fixed} number of quantifier alternations,
cannot be evaluated in polynomial sequential time by so-called
``branching-parsimonious algorithms''. This lower bound result
suggest that further research on the complexity of query
evaluation in constraint database theory should be directed
towards the identification of database and query classes that have
a strongly improved complexity behavior. As a pattern for the
development of such a theory, we suggest a new type of elimination
algorithms which are based on the notion of system degree and use
non-conventional data structures
(see~\cite{bghm1,bghm2,bghp,ghhmmp,ghmmp,gls,hkpsw,hmw,jkss,lecerf,schost}).

\bartq{remove this par?} This paper introduces a number of new
concepts for constraint database theory that sometimes require
certain notions from algebraic complexity theory, algebraic
geometry and commutative algebra. These notions can be found in
standard textbooks, such as \cite{bcs} (algebraic complexity
theory), \cite{atiyah} (commutative algebra) and
\cite{shafarevich} (algebraic geometry). The reader only
interested in database issues may read this paper while skipping
these technical details (and in particular the rather involved
proof of Theorems~\ref{lowerbound-geometric} and
~\ref{lowerbound-sample} below).

The remainder of this paper is organized as follows.

\section{Preliminaries on the constraint database model}\label{sec:cdbmodel}

  We define the notions of constraint database schema and instance.
 In Section \ref{}, we will discuss representations of database
instances. The definitions of constraint database schema and
instance are generalizations of the traditional definitions of
constraint databases \cite{cdbook} that also allow polynomial
functions to be included in a database.

 We
assume the existence of an infinite set of relation names and
function names.

\begin{definition}\label{def:dbschema}\rm
An \emph{constraint database schema}  is a finite sequence
$(R_1,...,R_r;F_1,...,F_s)$ of relation names $R_i$ ($i=1,...,r$)
and function names $F_j$ ($j=1,...,s$), where $r$ and $s$ are
integers. To each relation and function name $R_i$ and $F_j$,
natural numbers $ar(R_i)$ and $ar(F_j)$ are associated, called the
\emph{arity} of $R_i$ and $F_j$, respectively (we remark that
constants are modelled by  function names of arity zero).  \qed
\end{definition}

Let ${\cal S}=(R_1,...,R_r;F_1,...,F_s)$ be a constraint database
schema. Further on, we shall be interested in expressing queries
in first-order logic over the real numbers extended with the
relation and function names appearing in $\cal S$. We shall write
$\fopr{\cal S}$ for $\fopr{R_1,...,R_r;F_1,...,F_s}$ and
$\focr{\cal S}$  for $\focr{R_1,...,R_r;F_1,...,F_s}$ for these
 first-order languages with relation symbols
$R_1,...,R_r$ and function symbols $F_1,...,F_s$.

 We denote by $\R$ and $\C$ the sets of the real and
complex numbers.

\begin{definition}\label{def:dbinstance}\rm
Let ${\cal S}=(R_1,...,R_r;F_1,...,F_s)$ be a constraint database
schema. An \emph{constraint database instance} over $\cal S$ is a
finite sequence $(A_1,...,A_r;f_1,...,f_s)$ such that $A_i$ is a
semi-algebraic subset of $\R^{ar(R_i)}$ or a constructible
subset\footnote{Maybe Chevally's QE for algebraically closed
fields should be mentioned. The constraint database audience is
probably not familiar with that.} of  $\C^{ar(R_i)}$ ($i=1,...,r$)
and such that $f_j$ is a polynomial or rational function from
$\R^{ar(F_i)}$ to $\R$ or from $\C^{ar(F_i)}$ to $\C$
($j=1,...,s$).\qed
\end{definition}

 Below, we will refer to an
 constraint \emph{database schema} and \emph{instance}
simply as database schema and database instance.

Data objects such as the semi-algebraic and constructible sets and
polynomial functions can be modeled in various ways. In the
sequel, we shall use the term \emph{data model} to refer to a
conceptual model that is used to describe data objects.
Quantifier-free first-order formulas over the reals represent an
example of a data model which describes semi-algebraic sets. We
use the term \emph{data structure} to refer to the actual
structure that implements the data models. For example,
quantifier-free formulas may be given in disjunctive normal form
and the polynomials appearing in them may be given in dense or
sparse representation.

It is important to remark that there are different data structures
for constraint database instances and that for  each particular
structure  there are different representations of particular
instances. For example, relation and function instances can not
only be finitely described by means of (sometimes unique)
first-order formulas over the real numbers but also by, e.g.,
(\emph{never} unique) essentially division-free
\emph{arithmetic-boolean circuits} and \emph{arithmetic circuits},
respectively (see Section \ref{} and \ref{}).    For the purpose
of this section, it is enough to assume some fixed data structure.
When we speak about a \emph{representation} of a database
instance, it will be with respect to this fixed data structure.
The reader should think that representations of database instances
in this data structure are typically \emph{not} unique.

Next, we define the notion of query.

\begin{definition}\label{def:query}\rm
Let an input schema ${\cal S}=(R_1,...,R_r;F_1,...,F_s)$ and an
output schema $\tilde{\cal
S}=(\tilde{R}_1,...,\tilde{R}_p;\tilde{F}_1,...,\tilde{F}_q)$ be
given.

A \emph{query over the input schema ${\cal S}$ and output schema
$\tilde{\cal S}$} is a partial mapping $Q$  that maps database
instances over $\cal S$ to database instances over $\tilde{\cal
S}$. This mapping $Q$ can also be interpreted as a
 series of partial mappings
$(Q_{\tilde{R}_1},...,Q_{\tilde{R}_p};Q_{\tilde{F}_1},...,Q_{\tilde{F}_q})$,
where $Q_{\tilde{R}_i}$ ($i=1,...,p$)  maps database instances
over $\cal S$ to semi-algebraic or constructible subsets of
$\R^{ar(\tilde{R}_i)}$ or $\C^{ar(\tilde{R}_i)}$ and where
$Q_{\tilde{F}_j}$ ($j=1,...,q$) maps database instances over $\cal
S$ to a polynomial or rational function from
$\R^{ar(\tilde{F}_j)}$ to $\R$ or from $\C^{ar(\tilde{F}_j)}$ to
$\C$. We shall always suppose that queries
are induced by partial mappings which map representations of input
instances to representations of output instances. Hence,
semantically equivalent representations
 of input instances
(interpreted as semi-algebraic or constructible sets or polynomial
or rational functions) are mapped to semantically equivalent
representations of output instances.    These mappings of
representations of input instances or output instances always
will be given implicitly by the context and we shall not specify
them further. The reader may assume that these mappings are
computable (in some suitable sense), but we will not rely on this
fact in this paper.
 \qed
\end{definition}

We use the terminology \emph{relational query} and
\emph{functional query} in case of $q=0$ and $p=0$, respectively.

Below, we will simply refer to these mappings as \emph{queries}
whenever the input and output schemas  are clear.

We will be especially interested in relational and functional
queries that are expressible in extensions of first-order logic
over the real numbers, \fop, or complex number, \foc. Indeed,
given an input schema ${\cal S}=(R_1,...,R_r;F_1,...,F_s)$ and an
output schema $\tilde{\cal
S}=(\tilde{R}_1,...,\tilde{R}_p;\tilde{F}_1,...,\tilde{F}_q)$, we
can consider \fopr{{{\cal S},{\tilde{\cal S}}}} and \focr{{{\cal
S},{\tilde{\cal S}}}} as logics to express relations between   the
input and output databases  or as formalisms to specify relations
and functions. In particular, any formula $\varphi$ with
$k$ free variables in \fopr{{\cal S}
} or
\focr{{\cal S}
}, when evaluated on a database instance over ${\cal
S}=(R_1,...,R_r;F_1,...,F_s)$ defines a $k$-ary output relation,
when we interpret variables to range over the real or complex
numbers.

The formula $\varphi$ may be considered as a relational query that
corresponds to an output schema $\tilde{\cal S}$ which consist of
a single $k$-ary relation symbol.

\section{Geometric  elimination algorithms modeled as geometric queries}
 \label{sec:geometricelimination}

In this section, we argue by means of a number of examples of
elimination problems that geometric elimination algorithms can be
modeled as geometric queries (to be defined further in this
section) that satisfy some precise restrictions. We also discuss
the constraint formalism as a specification language.

\subsection{Elimination algorithms for non-parametric elimination problems}
\label{ssec:nonparametric}

As a first example, let us consider  the family (for varying
polynomial function $g$) of elimination problems

$$(\exists x_1)\cdots (\exists
x_n)(\bigwedge_{i=1}^{n}x_i^2-x_i=0\land
g(u_1,...,u_m,x_1,\ldots,x_n)=0),\eqno(\dagger_a)$$ in which
$u_1,...,u_m$ are considered parameters. Elimination problems of
the kind $(\dagger_a)$ are typically solved by algorithms that
produce as intermediate results suitable matrices
$M_{x_1},...,M_{x_n}\in \Q^{N\times N}$, which only depend on the
system $\wedge_{i=1}^{n}x_i^2-x_i=0$ (in case of the elimination
problem ($\dagger_a$), we have $N=2^n$). These matrices
$M_{x_1},...,M_{x_n}$ are defined in such a way that the
parameters $u_1,...,u_m$ satisfy the first-order formula
$(\dagger_a)$ if and only if they satisfy the quantifier-free
formula
$$\det{g(u_1,...,u_m,M_{x_1},...,M_{x_n})}=0.\eqno{(\dagger_b)}$$
By the way, let us observe that the formula $(\dagger_b)$
represents a declarative specification of the matrices
$M_{x_1},...,M_{x_n}$.

If the system $\wedge_{i=1}^{n}x_i^2-x_i=0$ is changed by one that
defines another zero-dimensional variety, in which polynomials of
degree at most $d$ appear, then $N$ will also change and in fact
be bounded by $d^n$. More generally, formulas of the form
$$(\exists x_1)\cdots (\exists x_n)(\varphi(x_1,...,x_n)\land
g(u_1,...,u_m,x_1,\ldots,x_n)=0), \eqno{(\dagger)}$$ where
$\varphi(x_1,...,x_n)$ defines a zero-dimensional $\Q$-definable
subvariety of $\C^n$, and $g$ is a polynomial function, form a
well-known class of elimination problems. Algorithms which solve
these elimination problems  produce typically matrices
 $M_{x_1},...,M_{x_n}$  as output. These matrices depend in size and
 content only on the subformula $\varphi(x_1,...,x_n)$ and they provide an
 easy way to obtain an elimination formula, namely $\det{g(u_1,...,u_m,M_{x_1},...,M_{x_n})}=0$,
for arbitrary $g$. These algorithms can therefore be seen as a
\emph{pre-processing} of the formula $\varphi(x_1,...,x_n)$ into
the matrices
  $M_{x_1},...,M_{x_n}$, aimed to facilitate the expression of the solution $\det{g(u_1,...,u_m,M_{x_1},...,M_{x_n})}=0$ of the solution of the elimination problem $(\dagger)$.

If we want to model this class of elimination problems and their
elimination polynomials using the language of constraint
databases, there are several ways of doing this. Here, we start
with  one simple formulation and then we explain how  the
formalism of constraint databases  can be  flexibly adapted to
deal with more general  situations.

For instance,   we may produce the elimination problem
$(\dagger_a)$ by applying the query expressed by the following
\focr{F_1,...,F_n,G}-formula
 $$(\exists
x_1)\cdots (\exists
x_n)(\bigwedge_{i=1}^{n}F_i(x_1,...,x_n)=0\land
G(u_1,...,u_m,x_1,\ldots,x_n)=0)\eqno(\dagger_c)$$ over the input
schema ${\cal S}=(F_1,...,F_n,G)$, to the database instance
$(f_1,...,f_n,g)$ given by $f_i(x_1,...,x_n):=x_i^2-x_i$
($i=1,...,n$) and some polynomial $g(u_1,...,u_m,x_1,...,x_n)$.

Similarly, the more general elimination problem $(\dagger)$ may be
produced by the query that is expressed by the \focr{R,G}-formula
$$(\exists x_1)\cdots (\exists x_n)(R(x_1,...,x_n)\land
G(u_1,...,u_m,x_1,\ldots,x_n)=0) $$
 over the input schema ${\cal S}'=(R,G)$, to the database instance $(A,g)$, where $A$ is supposed to be the zero-dimensional $\Q$-definable subvariety of $\C^n$ which is given by the \foc-formula $\varphi(x_1,...,x_n)$ and where $g(u_1,...,u_m,x_1,...,x_n)$ is some polynomial. In both cases, the output schema is $\tilde{\cal S}=\{\Delta\}$, where $\Delta$ is a function symbol, with $ar(\Delta)=m.$
 The database instance of $\tilde{\cal S}$ which represents the output of the query,
 is in both cases, the polynomial
 $\det{g(u_1,...,u_m,M_{x_1},...,M_{x_n})}.$

  Another way to produce the elimination problem $(\dagger)$ is the following: we suppose now that $g(u_1,...,u_m,x_1,...,x_n)$ is a fixed polynomial and we consider the query $Q_g$ expressed by the \focr{R}-formula
 $$(\exists
x_1)\cdots (\exists x_n)(R(x_1,...,x_n)\land
g(u_1,...,u_m,x_1,\ldots,x_n)=0) $$ over the input schema ${\cal
T}=(R)$.

We restrict our attention to input instances given by
\foc-formulas $\varphi(x_1,...,x_n)$ that define zero-dimensional
varieties of fixed (strictly positive) cardinality $N$.
Consequently, the output matrices $M_{x_1},...,M_{x_n}$ have also
fixed dimension $N$. In this way, we may view the output
$(M_{x_1},...,M_{x_n})$ of the underlying elimination
\bartq{$\ell$ should be $L$} algorithm as an instance of a schema
$\tilde{\cal T} =(\tilde{F}_1,...,\tilde{F}_\ell)$, with
$ar(\tilde{F}_j)=0$ ($j=1,...,\ell$), where $\ell=n\cdot N^2$.
Here, the function names $\tilde{F}_j$ ($j=1,..., \ell$) are
designed to describe  the constant entries of  the matrices
$M_{x_1},...,M_{x_n}$ (below, we will slightly abuse the notation
by regarding $(M_{x_1},...,M_{x_n})$ as an instance of the schema
$\tilde{\cal T} =(\tilde{F}_1,...,\tilde{F}_\ell)$). Observe that
this defines a query $Q$ with input schema ${\cal T}$ and output
schema $\tilde{\cal T}$ which maps the given $\Q$-definable input
instance $A\subset \C^n$ of cardinality $N$ to
$Q(A):=(M_{x_1},...,M_{x_n})$.

On the other hand, elimination problems like $(\dagger_b)$, are
obtained by applying the queries $\tilde{Q}_g$, expressed by
quantifier-free \focr{\tilde{F}_1,...,\tilde{F}_\ell}-formulas
$$\det g(u_1,...,u_m,\tilde{F}_1,...,\tilde{F}_\ell)=0 \eqno(\dagger_d)$$
over the input schema $\tilde{\cal S}
=(\tilde{F}_1,...,\tilde{F}_\ell)$ to the database instance
containing the $\ell=n\cdot N^2$ entries of the matrices
$M_{x_1},...,M_{x_n}$ as constants.

Hence, the original input database $A$ becomes pre-processed
 into a new databases consisting of the $n$-tuple of   matrices $(M_{x_1},...,M_{x_n})$ in such a way that, for arbitrary $g$, the query $Q_g$
applied to the input database $A$ and the query $\tilde{Q}_g$
applied to $(M_{x_1},...,M_{x_n})$ describe the same sets.
\ignore{Once the designer of the elimination algorithm has
described the translation of formula $(\dagger_c)$ into
$(\dagger_d)$, the  elimination algorithm itself can also be
viewed as a query $Q$ with input schema ${\cal S}=(F_1,...,F_n)$
and output schema $\tilde{\cal S}
=(\tilde{F}_1,...,\tilde{F}_\ell)$ that maps an input instance
$(f_1,...,f_n)$ over ${\cal S}$ to an output instance
$(M_{x_1},...,M_{x_n})$  over $\tilde{\cal S}$.} For any input
instance $A$, we have therefore
$$Q_g(A)=\tilde{Q}_g(Q(A)).$$
We observe that  the pre-processing performed by the underlying
elimination algorithm can be seen as the computation of a view on
the original database instance $A$, that allows the replacement of
the query $Q_g$, which is defined using quantifiers, by the query
$\tilde{Q}_g$, which can be expressed without quantifiers.

Below, we shall pay particular  attention to the following variant
of the family  of elimination problems $(\dagger_a)$, namely
$$(\exists x_1)\cdots (\exists
x_n)(\bigwedge_{i=1}^{n}x_i^2-x_i=0\land
y=g(u_1,...,u_m,x_1,\ldots,x_n)).\eqno(\ast_a)$$ Here,
$u_1,...,u_m$ are considered as parameters, $y$ as free variable
and  $g(u_1,...,u_m,x_1,...,x_n)$ is a suitable polynomial
function. Let  $M_{x_1},...,M_{x_n}\in \Q^{2^n\times 2^n}$ be the
matrices introduced before. Then all previous comments remain
valid, mutatis mutandis, if one replaces  the determinant in the
quantifier-free formula $(\dagger_b)$ by the characteristic
polynomial (in the free variable $y$), of the matrix
$$g(u_1,...,u_m,M_{x_1},...,M_{x_n}).$$

\subsection{Elimination algorithms for parametric elimination problems}
\label{ssec:parametric}

In this section, we discuss three variations of the previous
examples of elimination problem.

\subsubsection{First variation}\label{subsec:firstvariation}

We are going to consider the following variation of the example
$(\dagger)$ of Section~\ref{ssec:nonparametric}, where the
elimination problem takes the form
$$(\exists x_1)\cdots (\exists
x_n)(\bigwedge_{i=1}^n f_i(u_1,...,u_m,x_1,...,x_n)=0 \land
g(u_1,...,u_m,x_1,\ldots,x_n)=0). \eqno{(\dagger')}$$ Here
$u_1,...,u_m$ are parameters, possibly subject to some first-order
definable restriction (the parameters instances that satisfy this
restriction are called \emph{admissible\/} for the query
$(\dagger')$) and $f_1,...,f_n,g$ are polynomials in
$u_1,...,u_m,x_1,...,x_n$.

For fixed polynomial $g(u_1,...,u_m,x_1,...,x_n)$ the elimination
problem $(\dagger')$ can be produced by applying the query $Q'_g$,
expressed by the \focr{F_1,...,F_n}-formula
 $$(\exists
x_1)\cdots (\exists
x_n)(\bigwedge_{i=1}^{n}F_i(u_1,...,u_m,x_1,...,x_n)=0\land
g(u_1,...,u_m,x_1,\ldots,x_n)=0)\eqno(\dagger'_c)$$ over the input
schema ${\cal S}=(F_1,...,F_n)$, where now $ar(F_i)=m+n$
($i=1,...,n$), to some suitable database instance $(f_1,...,f_n)$.

For the sake of consistency and conciseness of exposition, we
shall require that the input database should satisfy the following
two \emph{flatness conditions}:
\begin{itemize}
\item the polynomials $f_1,...,f_n$ form a regular sequence in
$\Q[u_1,...,u_m,x_1,...,x_n]$; \item for
$V:=\{f_1=0,...,f_n=0\}\subset \C^m\times\C^n$, the morphism of
affine varieties $\pi:V\rightarrow \C^m$, induced by the canonical
projection $\C^m\times\C^n\rightarrow \C^m$, is finite.
\end{itemize}

The elimination problem $(\dagger')$ is typically solved by
algorithms that produce  matrices with \emph{rational} entries
$M_{x_1}(u_1,...,u_m),...,M_{x_n}(u_1,...,u_m)
 \in
\Q(u_1,...,u_m)^{N\times N}$.

We observe, that the flatness conditions satisfied by the input
database $(f_1,...,f_n)$ imply that the characteristic polynomial
(and in particular the determinant) of the matrix
$g(u_1,...,u_m,x_1,...,x_n)$ becomes a \emph{polynomial}
expression in $u_1,...,u_m$.

Therefore, if we wish to describe, in a similar way as in
Section~\ref{ssec:nonparametric}, such an  elimination algorithm
as a query mapping input database instances $(f_1,...,f_n)$ to
output database instances
$(M_{x_1}(u_1,...,u_m),...,M_{x_n}(u_1,...,u_m))$,  the constraint
database model should be extended and allow an output schema that
is not apriori fixed, e.g., using dynamic arrays of function names
in this case.

\ignore{A stronger flatness condition, expressing that the
$\wedge_{i=1}^{n}f_i(u_1,...,u_m,x_1,...,x_n)=0$ defines a
zero-dimensional variety of fixed cardinality could solve this
problem.}

Since our actual constraint database model is limited to fixed
schemas, we are obliged to require that the input database should
satisfy a third flatness condition, namely that the typical fiber
of the finite morphism of affine varieties $\pi:V\rightarrow \C^m$
should be of cardinality $N$, where $N$ is a previously fixed,
strictly positive integer.

Then we may fix an output schema $\tilde{\cal S}
=(\tilde{F}_1,...,\tilde{F}_\ell)$, as before, where $\ell=n\cdot
N^2$ and $ar(\tilde{F}_j)=m$ ($j=1,...,\ell$). An instance of
$\tilde{S}$ is now given by $\ell$ \emph{rational} functions
belonging to $\Q(u_1,...,u_m)$.

We call a parameter point $(\alpha_1,...,\alpha_m)\in \C^m$
admissible if the (rational) entries of
$M_{x_1}(u_1,...,u_m),...,M_{x_n}(u_1,...,u_m) $ are well-defined
in $\alpha_1,...,\alpha_m)$. Let us finally remark that the
entries of  the  output matrices
$(M_{x_1}(u_1,...,u_m),...,M_{x_n}(u_1,...,u_m))$ are not
arbitrary rational functions in $u_1,...,u_n$. Indeed, for every
input instance $(f_1,...,f_n)$ over ${\cal S}$, and every two
admissible parameter instances  $(\alpha_1,...,\alpha_m)$ and
$(\alpha'_1,...,\alpha'_m)$ for which $f_i( \alpha_1,\ab ...,\ab
\alpha_m,\ab x_1,\ab ...,\ab x_n)$ and $f_i( \alpha'_1,\ab ...,\ab
\alpha'_m,\ab x_1,\ab ...,\ab x_n)$ ($i=1,...,n$) are the same
polynomials (in $x_1,...,x_n$), also
$M_{x_j}(\alpha_1,...,\alpha_m)$ and
$M_{x_j}(\alpha'_1,...,\alpha'_m)$ will be equal. This remark will
turn out to be crucial for the lower bound results that follow in
the next section and motivate us to introduce the notion of
``geometric query'' at the end of this section.

\subsubsection{Second  variation}

Let $f_1,...,f_n$ be polynomials in the indeterminates
$u_1,...,u_m,x_1,...,x_n$ and assume that $f_1,...,f_n$ are
homogeneous of degrees $d_1,...,d_n$ with respect to
$x_1,...,x_n$. The elimination problem
$$(\exists
x_1)\cdots (\exists
x_n)(\bigwedge_{i=1}^{n}f_i(u_1,...,u_m,x_1,...,x_n)=0\land
\bigvee_{i=1}^n x_i\not= 0)$$ can be obtained by applying the
query, expressed by the \focr{F_1,...,F_n}-formula
$$(\exists
x_1)\cdots (\exists
x_n)(\bigwedge_{i=1}^{n}F_i(u_1,...,u_m,x_1,...,x_n)=0\land
\bigvee_{i=1}^n x_i\not= 0)$$ over the input schema ${\cal
S}=\{F_1,...,F_n\}$ with $ar(F_i)=m+n$ ($i=1,...,n$) to the
database instance $(f_1,...,f_n)$.
   Then
the output of an elimination algorithm could be the resultant  of
the polynomials $f_1,...,f_n$ with respect to $x_1,...,x_n$, which
we denote by denoted by ${\rm
Res}^{d_1,...,d_n}_{x_1,...,x_n}(f_1,...,f_n)$. Remark that
parametric systems of $n$  homogeneous linear equations in the
unknowns $x_1,...,x_n$ represent a particular case of this
situation. In this case the resultant becomes the determinant.

We observe now that  for every input instance $(f_1,...,f_n)$ over
${\cal S}=(F_1,...,F_n)$ and every two   parameter instances
$(\alpha_1,...,\alpha_m)$ and $(\alpha'_1,...,\alpha'_m)$ for
which $f_i( \alpha_1,\ab ...,\ab \alpha_m,\ab x_1,\ab ...,\ab
x_n)$ and $f_i( \alpha'_1,\ab ...,\ab \alpha'_m,\ab x_1,\ab
...,\ab x_n)$ ($i=1,...,n$) are the same polynomials, also ${\rm
Res}^{d_1,...,d_n}_{x_1,...,x_n}(f_1( \alpha_1,\ab ...,\ab
\alpha_m,\ab x_1,\ab ...,\ab x_n),...,f_n( \alpha_1,\ab ...,\ab
\alpha_m,\ab x_1,\ab ...,\ab x_n))$ and ${\rm
Res}^{d_1,...,d_n}_{x_1,...,x_n}(f_1( \alpha'_1,\ab ...,\ab
\alpha'_m,\ab x_1,\ab ...,\ab x_n),...,f_n( \alpha'_1,\ab ...,\ab
\alpha'_m,\ab x_1,\ab ...,\ab x_n))$ are the same polynomial.

\subsubsection{Third variation}  \label{subsec:thirdvariation}

As a last variation on of the example of
Section~\ref{ssec:nonparametric}, let us reconsider  the family
(for varying function $g$) of elimination problems
$$(\exists x_1)\cdots (\exists
x_n)(\bigwedge_{i=1}^{n}x_i^2-x_i=0\land
g(u_1,...,u_m,x_1,\ldots,x_n)=0),\eqno(\dagger_a)$$ where the part
$\wedge_{i=1}^{n}x_i^2-x_i=0$ is considered as fixed. This example
of a family of zero-dimensional elimination problems may seem
trivial, but in fact, no polynomial-time algorithm is known to
solve it. It turns out that this example already illustrates the
intrinsic  difficulty of geometric elimination.  For instance,
elimination algorithms which use simultaneous Newton iteration to
all zeroes of a zero-dimensional equation system lead to the
consideration of this kind of problem, see, e.g.,~\cite{}.

The complexity of the most efficient known elimination algorithms
for problem $(\dagger_a)$ depends exponentially on $n$, and
linearly on the circuit complexity of $g$.

The elimination problems $(\dagger_a)$ can be obtained by
applying the query expressed by the \focr{G}-formula
 $$(\exists
x_1)\cdots (\exists x_n)( \bigwedge_{i=1}^{n}x_i^2-x_i=0\land
G(u_1,...,u_m,x_1,\ldots,x_n)=0)\eqno(\dagger''_c)$$ over the
input schema ${\cal S}=(G)$, with $ar(G)=m+n$ , to the database
instance $(g)$ consisting of the polynomial function $g$. The
matrices $M_{x_1},...,M_{x_n}\in\Q^{2^n\times 2^n}$, introduced in
Section~\ref{ssec:nonparametric}, are well-defined  in this case
and the elimination algorithm under consideration produces as
output  a polynomial of the form
$\det{g(u_1,...,u_m,M_{x_1},...,M_{x_n})}.$ So, the elimination
algorithm can be modeled by a query with input schema ${\cal
S}=(G)$ and output schema $\tilde{\cal S}=(\tilde{G})$, where
$ar(\tilde{G})=m$.

Finally, we remark for any two admissible parameter instances
$(\alpha_1,...,\alpha_m)$ and $(\alpha'_1,...,\alpha'_m)$ for
which $g( \alpha_1,\ab ...,\ab \alpha_m,\ab x_1,\ab ...,\ab x_n)$
and $g( \alpha'_1,\ab ...,\ab \alpha'_m,\ab x_1,\ab ...,\ab x_n)$
($i=1,...,n$) are the same polynomials (in $x_1,...,x_n$), also
$\det{g(\alpha_1,...,\alpha_m,M_{x_1},...,M_{x_n})}$ and
$\det{g(\alpha'_1,...,\alpha'_m,M_{x_1},...,M_{x_n})}$ are the
same values.

\subsection{Definition of geometric queries}
\label{ssec:def-geometric}

The obserations made earlier in
Section~\ref{sec:geometricelimination} motivate the notion of
geometric query, which we shall introduce below. First, we shall
formalize a distinction between parameters and variables.

\subsubsection{Variables versus parameters}

From now on, we shall distinguish between two types of
indeterminates which we refer to as  \emph{variables} on the one
hand and as \emph{parameters} on the other hand.

 The difference will be reflected in the notation: we use
$u_1,u_2,\ldots$ to indicate \emph{parameters} and $x_1,x_2,\ldots
$ to indicate  \emph{variables}. In this section, we shall limit
our attention to this purely syntactic distinction between
variables and parameters and postpone the discussion of their
(difficult) semantic meaning to Section~\ref{sec:rep-model}.

In particular, when dealing with geometric elimination, we shall
always  work with an input database schema ${\cal
S}=(R_1,...,R_r;F_1,...,F_s)$ where all relation and function
symbols are of arity $m+n$ and which are such that the first $m$
indeterminates refer to parameters and the last $n$ to variables.
Our notation reflects this circumstance by by
$R_i(u_1,\ldots,u_m;x_1,\ldots,x_n)$ ($i=1,...,r$) and
$F_j(u_1,\ldots,u_m;x_1,\ldots,x_n)$ ($j=1,...,s$).

Let $A_i$ be instances of $R_i$ and $f_j$ be instances of $F_j$.
We shall always assume that there are \fop-formulas describing
 $A_i(u_1,\ldots,u_m;x_1,\ldots,x_n)$ and polynomials  (in the variables $x_1,...,x_m$) over the function field $\Q(u_1,...,u_m)$, describing
   $f_j(u_1,\ldots,u_m;x_1,\ldots,x_n)$ respectively.
For an admissable instance $(\alpha_1,...,\alpha_m)$ of the
parameters $u_1,...,u_m$, we shall always require that
$f_j(\alpha_1,...,\alpha_m,x_1,...,x_n)$ is a well-defined
polynomial in $x_1,...,x_n$.

\begin{definition}\label{def:eq-par}\rm
Let ${\cal S}=(R_1,...,R_r;F_1,...,F_s)$ be a database schema and
let
  $(A_1,...,A_r;f_1,...,f_s)$ be a database instance over ${\cal S}$.

  We call two admissible instances  $(\alpha_1,\ldots,\alpha_m)$ and
  $(\alpha'_1,\ldots,\alpha'_m)$ of  the parameters $u_1,...,u_m$
   \emph{equivalent} in the given database $(A_1,...,A_r;f_1,...,f_s)$ (or simply \emph{$(A_1,...,A_r;f_1,...,f_s)$-equivalent})
  if   the following conditions are satisfied:
  \begin{itemize}
  \item $\{(x_1,...,x_n)\in \R^n\mid
  A_i(\alpha_1,\ldots,\alpha_m,x_1,...,x_n)\}=\{(x_1,...,x_n)\in \R^n \mid
  A_i(\alpha'_1,\ldots,\alpha'_m,x_1,...,x_n)\}$ for all
  $i=1,...,r$; and
  \item
  $f_j(\alpha_1,\ldots,\alpha_m,x_1,...,x_n)$ and $f_j(\alpha'_1,\ldots,\alpha'_m,x_1,...,x_n)$
  are the same polynomials (in $x_1,...,x_n$) for all $j=1,...,s$.\qed
\end{itemize}
\end{definition}

\subsubsection{Definition of geometric queries}

\begin{definition}\label{def:geometricquery}\rm
Let  ${\cal S}=(R_1,...,R_r;F_1,...,F_s)$ be an input database
schema,  in which all relations and functions are of arity $m+n$
as discussed above, and let $\tilde{\cal
S}=(\tilde{R}_1,...,\tilde{R}_p;\tilde{F}_1,...,\tilde{F}_q)$ be
an output schema. For the sake of simplicity of exposition, we
assume that the relations and functions appearing in instances
over ${\cal S}$ and $\tilde{\cal S}$ depend on the same parameters
$u_1,...,u_m$.

A query
$Q=(Q_{\tilde{R}_1},...,Q_{\tilde{R}_p};Q_{\tilde{F}_1},...,Q_{\tilde{F}_q})
$ over the input schema ${\cal S}$ and output schema $\tilde{\cal
S}$ is called a \emph{geometric query} if for any database
instance
  $(A_1,...,A_r;f_1,...,f_s)$ and any  two admissible $(A_1,...,A_r;f_1,...,f_s)$-equivalent instances   $(\alpha_1,\ldots,\alpha_m)$ and
  $(\alpha'_1,\ldots,\alpha'_m)$ for the parameters $u_1,...,u_m$,
the parameter instances   $(\alpha_1,\ldots,\alpha_m)$ and
  $(\alpha'_1,\ldots,\alpha'_m)$
  are also admissible and equivalent in the database
  $$\displaylines{\quad Q(A_1,...,A_r;f_1,...,f_s)=\hfill{}\cr\hfill{}\quad (Q_{\tilde{R}_1}(A_1,...,A_r;f_1,...,f_s),...,Q_{\tilde{R}_p}(A_1,...,A_r;f_1,...,f_s);\hfill{}\cr\hfill{}Q_{\tilde{F}_1}(A_1,...,A_r;f_1,...,f_s),...,Q_{\tilde{F}_q}(A_1,...,A_r;f_1,...,f_s)).\quad \cr}$$
   \qed
\end{definition}

In the case of arbitrary first-order expressible queries, it may
occur that the parameters appearing in the relations and functions
in the
 output schema are different from those appearing in the relations and functions of the input schema.
Nevertheless, this fact does not really restrict the applicability
of our notion of  geometric query, since the formula that defines
the given first-order query may easily be rewritten (unnested)
into an equivalent one in which the input and output schema depend
on the same parameters. Since we will not relay on this more
general situation, we shall not go into more details.

It is clear that our notion of geometric query captures the
reality of existing elimination algorithms when the query-formula
only uses symbols belonging to the given database schema, as,
e.g., $(\dagger_c)$ in Section~\ref{ssec:nonparametric}. If the
query-formula contains both subexpressions without database
symbols and subexpressions with database sysmbols, as, e.g.,
$(\dagger'_c)$ in Section~\ref{subsec:firstvariation} and
$(\dagger''_c)$ in Section~\ref{subsec:thirdvariation}, then the
notion of geometric query still has an intuitive sense for
elimination algorithms. In the general case, the query-formula
must possibly be rewritten, before the formal definition of
geometric query may be meaningfully applied.

  We end this section with some trivial examples of geometric queries.
   The transformation that maps the function $u_1u_2x_1$ to the
function $x_1$ is an example of a geometric query, as is the
identity and the transformation that maps the functions
$u_1x_1,u_2x_1$ to $u_1u_2x_1$.

The transformation that maps the function $u_1u_2x_1$ to the
function $u_1$ is an example of a non-geometric query. Taking
$(\alpha_1, \alpha_2):=(1,0)$ and $(\alpha'_1,\alpha'_2)=(0,1)$
gives two equal functions for what concerns $u_1u_2x_1$, but not
for for what concerns $u_1$. \qed

\subsection{Specification of queries}

\subsubsection{First-order specification}

First-order logic over the real or complex numbers extended with
relation and function names, used to address input relations and
functions, allows to define output relations in the traditional
way.

As remarked in the introduction, the creation of output functions
requires an extension of first-order logic by special terms in
order to \emph{specify}  these functions. The
\fopr{F_{11},...,F_{nn},F}-formula $(\dagger)$ in the Introduction
is a first example of such a specification. Both the determinant
and the square of the determinant of any database instance of the
schema $(F_{11},...,F_{nn})$ satisfy this specification.

A less trivial example is given by the specification formula
 $$\displaylines{{\large (} \bigwedge_{i=1}^{n}(\forall
z)(\forall x_1)\cdots(\forall x_n)(F_i(u_1,...,u_m,x_1,...,x_n)=0
\hfill{}\cr\hfill{} \rightarrow F_i(u_1,...,u_m,zx_1,...,zx_n)=0)
\quad\cr\quad \land (\exists x_1)\cdots(\exists
x_n)(\bigwedge_{i=1}^{n}F_i(u_1,...,u_m,x_1,...,x_n)=0
\hfill{}(\dagger)\cr\hfill{}\land \bigvee_{i=1}^n x_i\not=0
){\large )}\leftrightarrow
\tilde{F}(u_1,...,u_m)=0\quad \cr}$$
over the input schema ${\cal S}=(F_1,...,F_n)$, with $ar(F_i)=m+n$
($i=1,...,n$) and output schema $\tilde{\cal S}=(\tilde{F})$ with
$ar(\tilde{F})=m$.

This formula can be interpreted as a descriptive specification of
a (typically geometric) query which maps instances $(f_1,...,f_n)$
over ${\cal S}=(F_1,...,F_n)$ to instances $(\tilde{f})$ of the
output schema $\tilde{\cal S}=(\tilde{F})$.

The reader should be aware that this formula, contrary to the
formula $(\dagger)$ of the Introduction, does not apriori restrict
the degree of the input polynomials with respect to the
$x_1,...,x_n$.

However, fixing an input $(f_1,...,f_n)$ that satisfies the
left-hand side of the equivalence in the specification
$(\dagger)$, the polynomials $f_i$ satisfy for suitable
non-negative integers $d_i$ the condition
$f_i(u_1,...,u_m,zx_1,...,zx_n)=z^{d_i}
f_i(u_1,...,u_m,x_1,...,x_n)$ ($i=1,...,n$). As we have already
observed in Section~\ref{subsec:firstvariation}, for each
occurrence of degrees $(d_1,...,d_n)$ the notion of resultant is
well defined and denoted by ${\rm
Res}^{d_1,...,d_n}_{x_1,...,x_n}(f_1,...,f_n)$. Hence if we
interpret the function symbol $\tilde{F}$ by ${\rm
Res}^{d_1,...,d_n}_{x_1,...,x_n}(f_1,...,f_n)$, we obtain the
output of a query which satisfies the given specification.

The traditional  informal specification of the resultant is
operative, but as illustrated here, the formalism of constraint
databases allows the declarative specification of objects such as
resultants and determinants.

\subsubsection{Higher-order specification}

Now, we consider an example of a specification of an elimination
task that cannot be formulated by means of a first-order formula.
We use a specification in English, based on database schemas as
before.

As, we have already seen in Section~\ref{ssec:nonparametric}
and~\ref{subsec:thirdvariation}, for varying polynomial function
$g(u_1,...,u_m,x_1,...,x_n)$, the family of elimination problems
expressed by $$(\exists x_1)\cdots (\exists
x_n)(\bigwedge_{i=1}^{n}x_i^2-x_i=0\land
y=g(u_1,...,u_m,x_1,\ldots,x_n)),\eqno(\ast_a)$$ can be
interpreted as the application of the query expressed by the
\fopr{G}-formula  $$(\exists x_1)\cdots (\exists
x_n)(\bigwedge_{i=1}^{n}x_i^2-x_i=0\land
y=G(u_1,...,u_m,x_1,\ldots,x_n))\eqno(\dagger''_c)$$ over the
input schema ${\cal S}=(G)$ to the database instance $(g)$
containing the polynomial function $g$.

The canonical elimination  polynomial of $(\ast_a)$ is
$$p(y,u_1,...,u_m):=\prod_{(\varepsilon_1,...,\varepsilon_n)\in \{0,1\}^n} (y-g(u_1,...,u_m, \varepsilon_1,...,\varepsilon_n)).\eqno(\ast_e)$$

The elimination algorithms, described in the geometric elimination
literature, aim at obtaining a representation of the polynomial
$p$, such that for any given values of the parameters
$u_1,...,u_m$, $p$ can be evaluated as a polynomial function in
$y$, e.g., by means of a division-free arithmetic circuit. More
specifically, the aim is to find  polynomials
$\omega_1(u_1,...,u_m), ...,\omega_\ell (u_1,...,u_m)$  and a
polynomial $$q(t_1,...,t_\ell ,y)=\sum_{0\leq j\leq 2^n}
g_j(t_1,...,t_\ell)y^j,$$ belonging to $\Q[t_1,...,t_\ell]$ and
$\Q[t_1,...,t_\ell,y]$ respectively, such that the identity

\ignore{\bigskip
\begin{tabular}{lll}
 $p(y,u_1,...,u_m)$&$=$&$q(\omega_1(u_1,...,u_m), ...,\omega_\ell (u_1,...,u_m),y)$\\
 &$=$&$\sum_{j=0}^{2^n}q_j(\omega_1(u_1,...,u_m), ...,\omega_\ell (u_1,...,u_m))y^j$
\end{tabular}
\bigskip}

$$\displaylines{\quad p(y,u_1,...,u_m)=q(\omega_1(u_1,...,u_m), ...,\omega_\ell (u_1,...,u_m),y)\hfill{}\cr\hfill{}=\sum_{j=0}^{2^n}q_j(\omega_1(u_1,...,u_m), ...,\omega_\ell (u_1,...,u_m))y^j\quad}$$
is satisfied.

The idea behind this elimination strategy is that of \emph{partial
evaluation}. This means that the evaluation of $p$ with respect to
$y$ is postponed and that $p(y,u_1,...,u_m)$ is  written, for some
fixed values of $u_1,...,u_m$, as a polynomial function $q$, in
some pre-processed parameter dependent values
$\omega_1(u_1,...,u_m),...,\omega_\ell(u_1,...,u_m)$ such that,
for the evaluation of $p$ in the input $y$, no branchings are
needed anymore.

 In the language of constraint databases, this pre-processing of $g(u_1,...,u_m,x_1,...,x_n)$ into the functions
 $\omega_1(u_1,...,u_m), ...,\omega_\ell (u_1,...,u_m)$, can be modeled as a query with input schema
 ${\cal
S}=(G)$  and output schema $\tilde{\cal
S}=(\tilde{\Omega_1},...,\tilde{\Omega_\ell })$, with
$ar(\Omega_j)=m$ ($j=1,...,\ell$).  We remark that, as soon as
$u_1,...,u_m$ and $\ell$ are fixed, also the database schema
$\tilde{\cal S}$ becomes fixed.

In the geometric elimination literature, algorithms are
(implicitly or explicitly) required to be  \emph{robust}
(see~\cite{cghmp} for the definition of the notion of robustness).
In the formalism of constraint databases, robustness can be
specified by requiring that the query, that transforms
$g(u_1,...,u_m,x_1,\ldots,x_n)$ into $\omega_1(u_1,...,u_m),
...,\omega_\ell (u_1,...,u_m)$, is a \emph{geometric query}.  This
means that for any two parameter instances
$(\alpha_1,...,\alpha_m)$ and $(\alpha'_1,...,\alpha'_m)$ for
which $g(\alpha_1,...,\alpha_m,x_1,\ldots,x_n)$ and
$g(\alpha'_1,...,\alpha'_m,x_1,\ldots,x_n)$ determine the same
polynomials in $x_1,...,x_n$, also
$\omega_1(\alpha_1,...,\alpha_m)=\ab
\omega_1(\alpha'_1,...,\alpha'_m),\ab  ...,\ab \omega_\ell
(\alpha_1,...,\alpha_m)=\ab \omega_\ell (\alpha'_1,...,\alpha'_m)$
holds. Therefore we may conclude that the query that transforms
the input database instance $(g)$ into the output database
instance $(p)$ is also a geometric query. On the other hand, we
observe that  $\omega(u_1,...,u_m),...,\omega_\ell(u_1,...u_m)$
may be interpreted as parameters of a fixed division-free circuit
with input $y$, which is represented by the polynomial $q$ and
which is independent of the input $g$. This circuit evaluates the
polynomial $p(u_1,...,u_m,y)$.

The task of transforming the elimination problem $(\ast_a)$, given
by the polynomial $g(u_1,...,u_m,x_1,...,x_n)$, into a collection
of intermediate polynomials $\omega_1(u_1,...,u_m),
...,\omega_\ell (u_1,...,u_m)$ that satisfy a number of
restrictions, can therefore be specified using the constraint
database formalism, as we have just illustrated.

We conclude this section by mentioning that we may   ask how long
the vector  $(\omega_1(u_1,...,u_m), ...,\omega_\ell
(u_1,...,u_m))$ has to be in order to be able to express
$p(u_1,...,u_m,y)$ by a fixed division-free arithmetic circuit
with input $y$, which is independent of the polynomial $g$. As the
main result of this paper, it turns out that \emph{the notion of
geometric query implies, that the number $\ell$, necessarily has
to be exponentially big in the number of quantified variables,
i.e., $\ell \geq 2^n$.}

\section{Representation of databases and algorithmic model for query evaluation}
\label{sec:rep-model}

 \subsection{Representation of databases}

The function symbols appearing in a database schema will be
interpreted by polynomial functions that are well-defined in any
commutative ring. Typically, the data structure to implement these
functions are division-free arithmetic circuits that allow the
evaluation of the function in an arbitrary commutative ring.

The relation symbols appearing in a database schema will be
interpreted by relations that may be evaluated in an arbitrary
commutative ring. The relations are
 implemented by
arithmetic boolean circuits.

Section~\ref{sec:cdbmodel}

These representations may be considered as implementations of a
certain datatype which allows in case of function symbols the
evaluation of the function in an arbitrary commutative ring and in
case of relation symbols the evaluation of membership

This paper uses notions from algebraic geometry and commutative
algebra. These notions can be found in standard textbooks, such as
\cite{shafarevich} and we refer to Appendix A for an overview of
them.

\subsection{Algorithmic models and complexity measures.}\label{subsec11}


\subsubsection{Data structures: essentially  division-free arithmetic circuits}

The algorithmic problems considered in this paper will depend on
{\em continuous parameters} and therefore the corresponding input
data structures have to contain entries for these parameters. We
call them {\em problem} or {\em input parameters}. Once such a
parametric problem is given, the specialization of the parameters
representing input objects are called (admissible) {\em problem}
or {\em input instances} and these may in principle be
algebraically dependent. An algorithm solving the given problem
operates on the corresponding input data structure and produces
for each admissible input instance an {\em output instance} which
belongs to a previously chosen output data structure.

The procedures or algorithms considered in this paper operate with
essentially  division-free arithmetic circuits as basic data
structures for the representation  of inputs and outputs. An {\em
essentially division--free arithmetic circuit}  (or straight-line
program) is an algorithmic device that can be represented by a
labeled directed acyclic graph ({\em dag}) as follows: the circuit
depends on certain input nodes, labeled by indeterminates over the
field of the rationals $\Q$. These indeterminates are thought to
be subdivided in two disjoints sets, representing the {\em
parameters} and the {\em variables} of the given circuit. For the
sake of definiteness, let $U_1\klk U_r$ be the parameters and
$Y_1\klk Y_t$ the variables of the circuit. Let $K:=\Q (U_1\klk
U_r)$ be the field of quotients of polynomials in $\Q [U_1\klk
U_r]$. We call $K$ the {\em parameter field} of the circuit. The
circuit nodes of indegree zero which are not inputs are labeled by
elements of $\Q$, which are called the {\em scalars} of the
circuit (here ``indegree" means the number of incoming edges of
the corresponding node). Internal nodes are labeled by arithmetic
operations (addition, subtraction, multiplication and division).
We require that the internal nodes of the circuit represent
polynomials in the variables $Y_1\klk Y_t$. We call these
polynomials the intermediate results of the given circuit. The
coefficients of these polynomials belong to the parameter field
$K$. In order to achieve this requirement, we allow in an
essentially division-free circuit only divisions which involve
elements of $K$. Thus essentially division--free circuits do not
contain divisions involving intermediate results which depend on
the variables $Y_1\klk Y_t$. A circuit which contains only
divisions by non--zero elements of $\Q$ is called {\em totally
division-free}.

Finally we suppose that the given circuit contains one or more
nodes which are labeled as output nodes. The results of these
nodes are called {\em outputs} of the circuit. Output nodes may
occur labeled additionally by sign marks of the form ``$=0$" or
``$\not=0$" \bartq{Also $>0$,$\geq 0$,...?} or may remain
unlabeled. Thus the given circuit represents by means of the
output nodes which are labeled by sign marks a system of
parametric polynomial equations and inequations. This system
determines in its turn for each admissible parameter instance a
locally closed set (i.e., an embedded affine variety) with respect
to the Zariski topology of the affine space $\A^t$ of variable
instances. The output nodes of the given circuit which remain
unlabeled by sign marks represent a parametric polynomial
application (in fact a morphism of algebraic varieties) which maps
for each admissible parameter instance the corresponding locally
closed set into a suitable affine space. We shall interpret the
system of polynomial equations and inequations represented by the
circuit as a {\em parametric family of systems} in the variables
of the circuit. The corresponding varieties constitute a {\em
parametric family of varieties}. The same point of view is applied
to the morphism determined by the unlabeled output nodes of the
circuit. We shall consider this morphism as a {\em parametric
family of morphisms}.\medskip

\subsubsection{Complexity models}
 To
a given essentially division-free arithmetic circuit we may
associate different complexity measures and models. In this paper
we shall be mainly concerned with {\em sequential} computing {\em
time}, measured by the {\em size} of the circuit. Occasionally we
will also refer to {\em parallel time}, measured by the {\em
depth} of the circuit. In our main complexity model is the {\em
total} one, where we take into account {\em all} arithmetic
operations (additions, subtractions, multiplications and possibly
occurring divisions) at {\em unit costs}. For purely technical
reasons we shall also consider two {\em non-scalar} complexity
models, one over the ground field $\Q$ and the other one over the
parameter field $K$. In the non-scalar complexity model over $K$
we count only the {\em essential} multiplications (i.e.
multiplications between intermediate results which actually
involve variables and not only parameters). This means that
$K$-linear operations (i.e. additions and multiplications by
arbitrary elements of $K$) are {\em cost free}. Similarly,
$\Q$-linear operations are not counted in the non-scalar model
over $\Q$.

Let $\theta_1\klk\theta_m$ be the elements of the parameter field
\bartq{Next 2 pars. needed?} $K$ computed by the given circuit.
Since this circuit is essentially division--free we conclude that
its outputs belong to $\Q[\theta_1\klk\theta_m][Y_1\klk Y_t]$. Let
$L$ be the non--scalar size (over $K$) of the given circuit and
suppose that  the circuit contains $q$ output nodes. Then the
circuit may be rearranged (without affecting its non--scalar
complexity nor its outputs) in such a way that the condition
\begin{equation}\label{arith_circuits} m=L^2+(2t-1)L+q(L+t+1)
\end{equation}
is satisfied. In the sequel we shall always assume that we have
already performed this rearrangement. Let \linebreak $Y:=(Y_1\klk
Y_t)$, $\theta:=(\theta_1\klk\theta_m)$ and let $f_1\klk f_q\in \Q
[\theta][Y]$ be the outputs of the given circuit. Let $Z_1\klk
Z_m$ be new indeterminates and write $Z:=(Z_1\klk Z_m)$. Then
there exist polynomials $F_1\klk F_q\in \Q [Z,Y]$ such that
$f_1=F_1(\theta,Y)\klk f_q=F_q(\theta,Y)$ holds. Let us write
$f:=(f_1\klk f_q)$ and $F:=(F_1\klk F_q)$. Consider the object
class
$$\mathcal{O}:=\{F(\zeta,Y):\zeta\in\A^m\}$$ which we think
represented by the data structure $\mathcal{D}:=\A^m$ by means of
the obvious encoding which maps each code $\zeta\in\mathcal{D}$ to
the object $F(\zeta,Y)\in\C[Y]^q$.

For the moment, let us consider as input data structure the
Zariski open subset $\mathcal{U}\subseteq \A^r$ where the rational
map $\theta=(\theta_1\klk\theta_m)$ is defined. Then the given
essentially division--free arithmetic circuit represents an
algorithm which computes for each input code $u\in\mathcal{U}$ an
output code $\theta(u)$ representing the output object
$f(u,Y)=F\big(\theta(u),Y\big)$. This algorithm is in the above
sense essentially division--free. From identity
(\ref{arith_circuits}) we deduce that the size $m$ of the data
structure $\mathcal{D}$ is closely related to the non--scalar size
$L$ of the given circuit. In particular we have the estimate
\begin{equation}\label{arith_circuits2} \sqrt{m}-(t+q)\le L.\end{equation}
Later we shall meet specific situations where we are able to
\bartq{Remove Later!?} deduce from a previous (mathematical)
knowledge of the mathematical object \linebreak $f=(f_1\klk f_q)$
a lower bound for the size of the output data structure of {\em
any} essentially division--free algorithm which computes for an
arbitrary input code $u\in\mathcal{U}$ the object $f(u,Y)$. Of
course, in such situations we obtain by means of
(\ref{arith_circuits2}) a lower bound for the non--scalar size
(over $K$) of {\em any} essentially division--free arithmetic
circuit which solves the same task. In particular we obtain lower
bounds for the total size and for the non--scalar size over $\Q$
of all such arithmetic circuits.

\subsubsection{Elimination problems}
 Given
an essentially division-free arithmetic circuit as input, an {\em
elimination problem} consists in the task of finding an
essentially division-free output circuit which describes the
Zariski closure of the image of the morphism determined by the
input circuit. The output circuit and the corresponding algebraic
variety are also called a {\em solution} of the given elimination
problem. We say that a given parameter point fixes an {\em
instance}  of the elimination problem under consideration. In this
sense a problem instance is described by an input and  an output
(or solution) instance.

    In this
paper we restrict our attention  to input circuits   which are
totally division-free and contain  only output nodes labelled by
``=0" and unlabelled output nodes. \bartq{Is this right?} Mostly
our output circuits will also be totally division-free and will
contain only one output node, labelled by the mark ``=0". This
output node will always represent a canonical elimination
polynomial associated to the elimination problem under
consideration (see Section \ref{subsec12} for more details).

 In case that
our output circuit contains divisions (depending only on
parameters but not on variables), we require to be able to perform
these divisions for any problem instance. In order to make this
requirement sound, we admit in our algorithmic model certain limit
processes in the spirit of de l'H{\^o}pital's rule (below we shall
modelise these limit processes algebraically, in terms of places
\bartq{Is technical def. necess.?} and valuations). The
restriction we impose on the possible divisions in an output
circuit represents a {\em first} fundamental geometric {\em
uniformity requirement\/} for our algorithmic model.

\subsubsection{Parametric elimination procedures}
 An algorithm which solves a given
elimination problem may be considered as a (geometric) elimination
procedure. However this simple minded notion is too restrictive
for our purpose of showing lower complexity bounds for elimination
problems. It is thinkable that there exists for every individual
elimination problem an efficient ad hoc algorithm, but that there
is no universal way to find and to represent all these ad hoc
procedures. Therefore, a {\em geometric elimination procedure\/}
in the sense of this paper will  satisfy certain uniformity and
universality requirements which we are going to explain now.

We modelise our elimination procedures by families  of {\em
arithmetic networks\/} (also called arithmetic-boolean circuits)
which solve entire\linebreak classes of elimination problems of
{\em arbitrary\/} input size. In this sense we shall require the
{\em universality\/} of our geometric elimination procedures.
Moreover, we require that our elimination procedures should be
essentially division-free.

 In a universal
geometric elimination procedure, branchings and divisions by
intermediate results (that involve only parameters, but not
variables) cannot be avoided. From our elimination procedures we
shall require to be {\em parsimonious} with respect to {\em
branchings} (and divisions). In particular we shall require that
our elimination procedures do not introduce branchings and
divisions for the solution of a given elimination problem when
traditional algorithms do not demand this (an example of such a
situation is given by the flat families of elimination problems we
are going to consider below in Section~\ref{subsec12}). This
restriction represents a {\em second} fundamental {\em uniformity
requirement} for our algorithmic model.

We call a universal elimination procedure {\em parametric\/} if it
satisfies our first and second uniformity requirement, i.e., if
the procedure does not contain
 branchings which otherwise could be avoided and if all possibly occurring
divisions can be performed on all problem instances, in the way we
have explained before. In this paper we shall only consider
parametric elimination procedures.

We call a parametric elimination procedure {\em geometrically
robust}  if it produces for any input instance  an output  circuit
which depends only on the mathematical objects ``input equation
system'' and  ``input   morphism'' but not on their circuit
representation. We shall apply this notion only to elimination
problems given by (geometrically or scheme-theoretically) flat
families of algebraic varieties. This means informally that a
parametric elimination procedure is geometrically robust if it
produces  for  flat families of problem  instances ``continuous''
solutions.

 Of course,
our notion of geometric robustness depends on the (geometric or
scheme-theoretical) context, i.e. it is not the same for schemes
or varieties. In Section \ref{subsec12}  we shall explain our idea
of geometric robustness   in the typical situation of flat
families of algebraic varieties given by reduced complete
intersections.

  Traditionally, the
size of a system of polynomial equations (and inequations) is
measured in purely extrinsic, syntactic terms (e.g. number of
parameters and variables, degree of the input polynomials, size
and depth of the input circuit etc). However, there exists a new
generation of symbolic and numeric algorithms which take also into
account intrinsic, semantic (e.g. geometric or arithmetic)
invariants of the input equation system in order to measure the
complexity of elimination procedure under consideration more
accurately.

 In this paper we shall turn back to the \bartq{This par. remains?}
traditional
 point of view. In \cite{cghmp} it was shown  that, under certain
universality and
 uniformity restrictions, {\em no parametric elimination
procedure which includes efficient computation of Zariski closures
and of generically squarefree parametric greatest common divisors
for circuit represented algebraic families of polynomials, is able
to solve  an arbitrary elimination problem in polynomial
(sequential) time, if time is  measured in terms of circuit size
and input length is measured in  syntactical
 terms only}.

\subsection{Flat families of elimination problems}\label{subsec12}

\subsubsection{Definitions and preliminaries}

Let, as before, let $U_1,\ldots, U_r, X_1,\ldots, X_n, Y$ be
indeterminates over $\Q$. In the sequel we shall consider
$X_1,\ldots, X_n$ and $Y$ as variables and $U_1,\ldots, U_r$ as
parameters. Let $G_1,\ldots, G_n$ and $F$ be polynomials belonging
to the $\Q$-algebra $\Q[U_1,\ldots, U_r, X_1,\ldots, X_n]$.
Suppose that the polynomials $G_1,\ldots, G_n$ form a regular
sequence \bartq{regular seq.?} in
 $\Q[U_1,\ldots, U_r, X_1,\ldots, X_n]$ defining thus an equidimensional
 subvariety $V:=\{
G_1=0\klk G_n=0\}$ of the $(r+n)$-dimensional affine space $\A^{r}
\times \A^{n}$ over the field $\C$. The algebraic variety $V$ has
dimension $r$. Let $\delta$ be the (geometric) degree of $V$ (this
degree does not take
 into account multiplicities or components at infinity). Suppose
furthermore that the morphism of affine varieties
$\pi:V\longrightarrow \A^{r}$, induced by the canonical projection
of $ \A^{r} \times \A^{n}$ onto $ \A^{r}$, is finite and
generically unramified (this implies that $\pi$ is flat and that
the ideal generated by $G_1,\ldots, G_n$ in $\Q [U_1,\ldots, U_r,
X_1,\ldots, X_n]$ is radical). Let $\tilde{\pi}:V\longrightarrow
\A^{r+1}$ be the morphism defined by $\tilde{\pi}(z) :=
(\pi(z),F(z))$ for any point $z$ of the variety $V$. The image of
$\tilde{\pi}$ is a hypersurface of $\A^{r+1}$ whose minimal
equation is a polynomial of $ k[U_1,\ldots, U_r, Y]$ which we
denote by $P$. Let us write $\deg P$ for the total degree of the
polynomial $P$ and $\deg_Y P$ for its partial degree in the
variable $Y$. Observe that $P$ is monic in $Y$ and that $\deg
P\leq \delta \deg F$ holds. Furthermore, for a Zariski dense set
of points $u$ of $\A^{r}$, we have that $\deg _Y P$ is the
cardinality of the image of the restriction of $F$ to the finite
set $\pi^{-1}(u)$. The polynomial $P(U_1,\ldots, U_r,F)$ vanishes
on the variety $V$.

Let us consider an arbitrary point $u=(u_1,\ldots ,u_r)$ of $\A
^r$. For arbitrary polynomials \scalebox{.978}[1]{$A\in \Q
[U_1,\ldots, U_r,X_1,\ldots, X_n]$} and \scalebox{.978}[1]{$B\in
\Q[U_1,\ldots, U_r,Y]$} we denote by $A^{(u)}$ and $B^{(u)}$ the
polynomials $A(u_1\klk u_r,X_1,\ldots, X_n)$ and $B(u_1\klk
u_r,Y)$ which belong to $\Q (u_1\klk u_r)[X_1,\ldots, X_n]$
and\linebreak $\Q (u_1\klk u_r)[Y]$ respectively. Similarly we
denote for an arbitrary polynomial $C\in \Q [U_1,\ldots, U_r]$ by
$C^{(u)}$ the value $C(u_1,\ldots ,u_r)$ which belongs to the
field $\Q (u_1,\ldots ,u_r)$. The polynomials $G_1^{(u)}\klk
G_n^{(u)}$ define a zero dimensional subvariety $V^{(u)}:=
\{G_1^{(u)}=0\klk G_n^{(u)}=0\}= \pi^{-1}(u)$ of the affine space
$\A^{n}$. The degree (cardinality) of $V^{(u)}$ is bounded by
$\delta$. Denote by  $\tilde{\pi}^{(u)}:\ V^{(u)} \longrightarrow
\A^{1}$ the morphisms induced by the polynomial $F^{(u)}$ on the
variety $V^{(u)}$. Observe that the polynomial $P^{(u)}$ vanishes
on the (finite) image of the morphism $\tilde{\pi}^{(u)}$. Observe
also that the polynomial $P^{(u)}$ is not necessarily the minimal
equation of the image of $\tilde{\pi}^{(u)}$.

We call the equation system $G_1 =0\klk G_n =0$ and the polynomial
$F$ {\em a flat family of elimination problems depending on the
parameters $U_1,\ldots, U_r$} and we call $P$ the associated {\em
elimination polynomial}. An element $u\in\A^r$ is considered as a
{\em parameter point} which determines a {\em particular problem
instance} (see Section \ref{subsec11}). The equation system $G_1
=0\klk G_n =0$ together with the polynomial $F$ is called the {\em
general instance} of the given flat family of elimination problems
and the elimination polynomial $P$ is called the {\em general
solution} of this flat family.

The {\em problem instance} determined by the parameter point $u\in
\A^r$ is given by the equations $G_1^{(u)}=0\klk G_n^{(u)}=0$ and
the polynomial $F^{(u)}$. The polynomial $P^{(u)}$ is called {\em
a solution} of this particular problem instance. We call two
parameter points $u,u' \in \A^r$ {\em equivalent} (in symbols: $u
\sim u'$) if $G_1^{(u)}=G_1^{(u')} \klk G_n^{(u)}=G_n^{(u')}$ and
$F^{(u)} = F^{(u')}$ holds. Observe that $u \sim u'$ implies
$P^{(u)} = P^{(u')}$. We call polynomials $A\in \Q [U_1,\ldots,
U_r,X_1,\ldots, X_n]$, $B\in \Q [U_1,\ldots, U_r,Y]$ and $C\in \Q
[U_1,\ldots, U_r]$ {\em invariant} (with respect to $\sim$) if for
any two parameter points $u,u'$ of $\A^r$ with $u\sim u'$ the
respective identities $A^{(u)} = A^{(u')}$, $B^{(u)} = B^{(u')}$
and $C^{(u)}=C^{(u')}$ hold.

\subsubsection{Arithmetic circuits}

An arithmetic circuit in $\Q [U_1,\ldots ,U_r,Y]$ {\em with
scalars in $\Q [U_1,\ldots ,U_r]$} is a totally division-free
arithmetic circuit in $\Q [U_1,\ldots ,U_r,Y]$, say $\beta$,
modelised in the following way: $\beta$ is given by a directed
acyclic graph whose internal nodes are labelled as before by
arithmetic operations. There is only one input node of $\beta$,
labelled by the variable $Y$. The other nodes of indegree zero the
circuit $\beta$ may contain, are labelled by arbitrary elements of
$\Q [U_1,\ldots ,U_r]$. These elements are considered as the {\em
scalars of $\beta$\/}. We call such an arithmetic circuit $\beta$
{\em invariant } (with respect to the equivalence relation $\sim
$) if all its scalars are invariant polynomials of $\Q
[U_1,\ldots ,U_r]$. Considering instead of $Y$  the variables
$X_1,\ldots ,X_n$ as inputs, one may analogously define the notion
of an arithmetic circuit in $\Q [U_1,\ldots ,U_r,X_1,\ldots ,X_n]$
with scalars in $\Q  [U_1,\ldots ,U_r]$ and the meaning of its
invariance. However, typically  we shall limit ourselves to
circuits in $\Q  [U_1,\ldots,U_r,Y]$ with scalars in $\Q
[U_1,\ldots ,U_r]$.

\subsubsection{Geometrically robust parametric elimination problems}

We are now ready to characterise in the given situation what we
mean by a {\em geometrically robust parametric elimination
procedure}. \bartq{Robust more tech. [CGHMP].} Suppose that the
polynomials $G_1,\ldots, G_n$ and $F$ are given by a totally
division-free arithmetic circuit $\beta$ in $\Q [U_1,\ldots, U_r,
X_1,\ldots, X_n]$. A geometrically robust parametric elimination
procedure accepts the circuit $\beta$ as input and produces as
output an {\sl invariant} circuit $\Gamma$ in $\Q [U_1,\ldots,
U_r, Y]$ with scalars in $\Q [U_1,\ldots, U_r]$, such that
$\Gamma$ represents the polynomial $P$. Observe that in our
definition of geometric robustness we did not require that $\beta$
is an invariant circuit because this would be too restrictive for
the modelling of concrete situations in computational elimination
theory.

The invariance property required for the output circuit $\Gamma$
means the following: let $u=(u_1\klk u_r)$ be a parameter point of
$\A^r$ and let $\Gamma^{(u)}$ be the arithmetic circuit in $\Q
(u_1\klk u_r)[Y]$ obtained from the circuit $\Gamma$ evaluating in
the point $u$ the elements of $\Q [U_1,\ldots ,U_r]$ which occur
as scalars of $\Gamma$. Then the invariance of $\Gamma$ means that
the circuit $\Gamma^{(u)}$ depends only on the particular {\em
problem instance} determined by the parameter point $u$ but not on
$u$ itself. Said otherwise, a geometrically robust elimination
procedure produces the solution of a particular problem instance
in a way which is independent of the possibly different
representations of the given problem instance.

By definition, a geometrically robust parametric elimination
procedure produces always the {\em general} solution of the flat
family of elimination problems under consideration. This means
that for flat families, geometrically robust parametric
elimination procedures do not introduce branchings in the output
circuits. It turns out that the following meta-statement becomes
true:
 {\em within the standard philosophy
  of commutative algebra, none of
  the known (exponential time) parametric elimination procedures can be
  improved to a polynomial time algorithm.\/} For this purpose it is
important to remark that the known {\em parametric\/} elimination
procedures (which are without exception based on linear algebra as
well as on comprehensive Gr{\"o}bner basis techniques) are all
geometrically robust.

The invariance property of these procedures is easily verified in
the situation of a flat family of elimination problems. One has
only
 to observe that all known elimination
procedures accept the input polynomials $G_1,\ldots, G_n$ and $F$
in their dense or sparse {\em coefficient representation\/} or as
{\em evaluation black box\/} with respect to the variables
$X_1,\ldots, X_n$.

Finally let us observe that robust elimination procedures can be
specified as geometric queries.

\section{A lower complexity bound for evaluating geometric queries}

As main result of this paper, we obtain that \emph{for geometric
FO - queries the size of the output data schema may become
necessarily exponentially big in the number of quantified
variables occurring in the query.}

The proof of this fact goes along the lines of Theorem
\label{lowerbound} below.

\section{Sample point queries}

\subsection{Sample point
queries and generalized sample point queries}

\subsubsection{The example of rationally parameterized families of
polynomial functions.} A particular instance of interest is the
case that the semi-algebraic set $A$ is contained in $\R^{m+n+1}$
and represents a rational family of polynomial functions from
$\R^n$ to $\R$. To be more precise, let $\pi:\R^{m+n+1}\rightarrow
\R^m$ be the canonical projection of any point of $\R^{m+n+1}$ on
its first $m$ coordinates. Suppose that $A$ is non-empty and that
for any $u=(u_1,\ab \ldots, \ab u_m) \in \pi(A)$ the
semi-algebraic set $(\{u\}\times \R^{n+1})\cap A$ is the graph of
an $n$-variate polynomial $f_u\in \R[x_1,\ldots,x_n]$. It is a
natural extension of our previously introduced sample-point query
to ask for a procedure which enables us for each $u\in \pi(A)$ and
each $x\in\R^n$ to compute the value of $f_u(x)$. The output of
such a procedure may be a purely existential prenex first-order
formula in the free variables $u_1,\ab \ldots, \ab u_m$ and
$x_1,\ab \ldots, \ab x_n$ which represents for each $u\in \pi(A)$
a division-free arithmetic circuit which evaluates the polynomial
$f_u$ (observe that there exists a uniform degree bound for all
 these polynomials). One easily verifies that all our requirements on the
semi-algebraic set $A$, except that of the polynomial character of
the function represented by the graph $(\{u\}\times \R^{n+1})\cap
A$, are first-order definable over the reals. Nevertheless, over
the complex numbers, when $A$ is a constructible (i.e., a
first-order definable subset of $\C^{m+n+1}$), all these
requirements are first-order expressible. This leads us to a new
type of computable queries which return on input a semi-algebraic
or constructible set $A$ as above and an element $u\in \pi(A)$, a
first-order formula which represents a division-free arithmetic
circuit evaluating the polynomial $f_u$. Uniformity of query
evaluation with respect to $u$ is expressed by the requirement
that the terms contained in this formula have to depend
\emph{rationally} on $u$.

In the following, we shall refer to this type of queries as
\emph{extended sample point queries}. We shall refer to
$u_1,\ldots,u_m$ as the \emph{parameters} and to $x_1,\ldots,x_n$
as the \emph{variables} of the query.

Suppose now that $A$ is a constructible subset of $\C^{m+n+1}$
with irreducible Zariski-closure. Let $V$ be the Zariski-closure
of $\pi(A)$. Then $V$ is an irreducible affine subvariety of
$\C^m$. We denote the function field of $V$ by $K$. It is not
difficult to see that for generically chosen parameter points
$u\in \pi(A)$, the extended sample point query associated to $A$
can be realized by a greatest common divisor computation in the
polynomial ring $K[x_1,\ldots,x_n]$.

\subsubsection{Variables versus parameters.}
The previous example is motivated by spatial data that come from
physical observation and are only known with uncertainty. Another
motivation comes from parametric optimization. Optimization
problems described in can also be studied in parametric form,
i.e., in the case where the linear inequalities and the target
function contain coefficients that depend on a time parameter
\cite{bank} and \cite{heku}. In this case, an optimum is not an
arbitrary set of sample points but an analytic (or at least
continuous) function which depends on a time parameter.

We are now going to explain why we distinguished between the
parameters $u_1,\ldots,u_m$ and the variables  $x_1,\ab \ldots,
\ab x_n$ in our discussion of rationally parameterized families of
polynomial functions. In the example above, let $\Phi(u_1,\ab
\ldots, \ab u_m;x_1,\ldots, x_m,y)$ be a quantifier-free formula
which defines the semi-algebraic or constructible set $A$. Let us
suppose that  $\Phi$ contains a subformula $\Psi(u_1,\ab \ldots,
\ab u_m)$ which expresses an internal algebraic dependency between
the parameters $u_1,\ab \ldots, \ab u_m$.  With respect to the
variables $x_1,\ab \ldots, \ab x_n$ there is no such subformula
contained in $\Phi$. For the sake of simplicity, we shall suppose
that $\Psi$ defines the set $\pi(A)$ and that there exists a
formula $\Omega(u_1,\ab \ldots, \ab u_m;x_1,\ab \ldots, \ab
x_n,y)$ such that $\Phi(u_1,\ab \ldots, \ab u_m;x_1,\ldots,
x_m,y)$ can be written as $$\Psi(u_1,\ab \ldots, \ab u_m) \land
\Omega(u_1,\ab \ldots, \ab u_m;x_1,\ldots, x_m,y).$$

Below we shall meet natural examples of parameterized algebraic
families of polynomial functions where the formula $\Psi$ becomes
of uncontrolled size and is of few interest, whereas the formula
$\Omega$ becomes the relevant part of the output information of a
suitable elimination algorithm. This situation occurs for instance
when the points $u$ of $\R^m$ satisfying the formula $\Psi$ are
given in parametric form (i.e., when they are image points of some
polynomial or rational map coming from some affine source space).
In this case, we are only interested in the subformula $\Omega$,
since points $u\in \R^m$ satisfying $\Psi$ can easily be produced
in sufficient quantity. In subsequent queries, $x_1,\ldots,x_n$
may appear as bounded variables, whereas the parameters
$u_1,\ldots,u_m$ are not supposed to be subject to quantification.
The example of the $u_1,\ldots,u_m$ expressing uncertainty in
physical spatial data illustrates this. These different r\^oles
motivate us to distinguish between $u_1,\ldots,u_m$ and $x_1,\ab
\ldots, \ab x_n$ and to call them parameters and variables,
respectively.

\subsubsection{The branching-parsimonious algorithmic model.}
In the model, that we are going to use in the sequel, parameters
and variables receive a different treatment.  (Free) variables may
be specialized into arbitrary real (or complex) values, whereas
the specialization of parameters may be subject to certain
restrictions. In the above example of a rationally parameterized
family of polynomial functions, the $n$-tuple of variables
$(x_1,\ldots,x_n)$ may be specialized into any point of the affine
space $\R^n$ (or $\C^n$), whereas the $m$-tuple of parameters
$(u_1,\ab \ldots, \ab u_m)$ may only be specialized into points
satisfying $\Psi(u_1,\ab \ldots, \ab u_m)$, i.e., into points
belonging to $\pi(A)$. Once the $n$-tuple of variables
$(x_1,\ldots,x_n)$ is specialized into a point of the
corresponding affine space, this point cannot be modified anymore.
However, we allow infinitesimal modifications of a given
specialization of the $m$-tuple of parameters $(u_1,\ab \ldots,
\ab u_m)$ within the domain of definition determined by the
formula $\Psi$. In the branching-parsimonious model, we require
that an arithmetic boolean circuit which represents the
semi-algebraic or constructible set $A$ does not contain divisions
which involve the variables $x_1,\ab \ldots, \ab x_n$. Similarly,
for a given point $u\in \pi(A)$, we require that the arithmetic
circuit representing the polynomial $f_u$ is division-free.
However, divisions by algebraic expressions in the parameters
$u_1,\ab \ldots, \ab u_m$ are sometimes unavoidable (e.g., in the
case of parametric greatest common divisor computations;
see~\cite{cghmp,gihe}). Therefore, we allow certain limited
divisions by algebraic expressions which depend only on the
parameters $u_1,\ab \ldots, \ab u_m$. More precisely, we allow
that the arithmetic boolean circuits representing the set $A$ or
the output of the corresponding extended sample point query
computes certain, but not arbitrary,
 rational functions
in the parameters $u_1,\ab \ldots, \ab u_m$,  called
\emph{scalars} of the circuit. However, we do not allow the
division of a positive-degree polynomial in the variables $x_1,\ab
\ldots, \ab x_n$ by a non-constant scalar. In the above sense, we
require for our branching-parsimonious algorithmic model that
arithmetic boolean circuits are \emph{essentially division-free
with respect to variables} (see~\cite{cghmp,gihe}  for a precise
definition).

\subsubsection{Branching-free output representations of
extended sample point queries.} Since we allow certain
infinitesimal modifications of the parameters $u_1,\ab \ldots, \ab
u_m$ within their domain of definition, we sometimes may replace
divisions (and corresponding branchings) by limit processes in the
spirit of L'H\^{o}pital's rule. It is possible to mimic
algebraically this kind of limit process by places
(see~\cite{lang} for the notion of place and~\cite{cghmp,gihe} for
motivations of this idea).

Branchings corresponding to divisions can   trivially be avoided
by restricting input data. Therefore a meaningful notion of
branching-parsimonious (or branching-free) algorithm requires the
consideration of Zariski-closures of input data sets. This may
partially explain the rather technical assumptions and tools in
the following ad hoc definition of a branching-free representation
of the output of an extended sample point query.

Suppose now that in the example above $A$ is a constructible
subset of $\C^{m+n+1}$ with irreducible Zariski-closure $B$. Let
$V$ be the Zariski-closure of $\pi(A)$ in $\C^{m}$. Then $V$ is an
irreducible affine variety whose function field we denote by $K$.
Moreover, the irreducible affine variety $B$ is birationally
equivalent to $V\times \C^n$. Suppose furthermore that $\pi(B)=V$
holds and that $B$ represents a rationally parameterized family of
polynomial functions which extends the family represented by $A$.
Then we say that the extended sample point query associated with
$A$ admits a \emph{branching-free output representation} if there
exists  an essentially division-free, single-output arithmetic
circuit $\beta$ with inputs $x_1, \ldots, x_n$ and scalars
$\theta_1, \ldots, \theta_s\in K$ satisfying the following
conditions:

\begin{enumerate}
\item[(i)] for any point $u\in\pi(A)$ where the rational functions
$\theta_1, \ldots, \theta_s$ are defined, the division-free
arithmetic circuit, obtained from $\beta$ by specializing the
scalars $\theta_1, \ldots, \theta_s$ into the complex values
$\theta_1(u), \ldots, \theta_s (u)$, evaluates the polynomial
$f_u$;

\item[(ii)] for any point $u\in V$ and any place
$\varphi:K\rightarrow \C\cup \{\infty\}$ whose valuation ring
extends the local ring of the affine variety $V$ at the point $u$,
the values $\varphi(\theta_1) ,\ldots, \varphi(\theta_s)$ are
finite and uniquely determined  by $u$ (therefore we shall write
$\theta_1(u):=\varphi(\theta_1),\ldots ,
\theta_s(u):=\varphi(\theta_s)$).
\end{enumerate}

Let an arithmetic circuit $\beta$ be given as above. Then we call
$\beta$ a branching-free representation of the output of the
extended sample point query associated to $A$.

Observe that the output of the circuit $\beta$ represents a
polynomial belonging to $K[x_1,\ldots, x_n]$ whose coefficients
satisfy condition (ii). Moreover, the arithmetic circuit $\beta$
constitutes a division-free representation of the extended sample
point query associated to the Zariski-closure $B$ of $A$. Finally,
let us remark that for any $u\in V$, $x\in \C^n$ and $y\in \C$,
the point $(u,x,y)$ belongs to $B$ if and only if the circuit
$\beta_u$, obtained from $\beta$ by replacing the scalars
$\theta_1,\ldots ,\theta_s$ by the complex numbers
$\theta_1(u),\ldots ,\theta_s(u)$, computes on input $x$ the
output $y$.

We require that a \emph{branching-parsimonious query evaluation
algorithm} produces a branching-free output representation of the
given extended sample point query if the query admits such a
representation.

Let us also observe that extended sample point queries appear in a
natural way if we apply the constraint database concept to data
processing in the context of approximation theory and functional
analysis.

\subsection{A lower complexity bound for evaluating sample point
queries}

In this section, we restrict our attention to constraint databases
defined in the language \foeq\ over the complex numbers. We shall
consider two ternary relational predicates, namely $S(v_1,v_2,w)$
and $P(v_1,v_2,w)$. Our query language will therefore be
$\mathrm{FO}(+,\times,0,1,=,S,P)$. Let $L,n$ be given natural
numbers and let $r:=(L+n+1)^2$. For any polynomial $f\in
\C[x_1,\ldots,x_n]$, we denote by $L(f)$ the minimal non-scalar
size of all division-free arithmetic circuits with inputs
$x_1,\ldots,x_n$ and scalars from $\C$ which evaluate the
polynomial $f$. Let $$W_{L,n}:=\{f\in \C[x_1,\ldots,x_n]\mid
L(f)\leq L\}.$$ One sees easily that all polynomials contained in
$W_{L,n}$ have degree at most $2^L$ and that $W_{L,n}$ forms a
$\Q$-definable object class which has a $\Q$-definable holomorphic
encoding by the continuous data structure $\C^r$. Observe that
Zariski-closure $\overline{W_{L,n}}$ of $W_{L,n}$ is a
$\Q$-definable, absolutely irreducible algebraic variety
consisting of the polynomials of $\C[x_1,\ldots,x_n]$ which have
approximate complexity at most $L$. Moreover, the affine variety
$\overline{W_{L,n}}$ forms a cone in its ambient space (i.e., for
any $\lambda\in \C$ we have $\lambda \overline{W_{L,n}}\subseteq
\overline{W_{L,n}}$). For details on complexity and data structure
models we refer to \cite{bcs,cghmp}.

Let $z_1, \ldots, z_r$ and $y$ be new variables. Choose now a
directed acyclic graph ${\cal D}_{L,n}$ representing a generic,
division-free arithmetic circuit with input nodes $x_1,\ldots,
x_n$, output node $y$ and scalar nodes $z_1,\ldots, z_r$ such that
any polynomial of $W_{L,n}$ may be evaluated by the division-free
arithmetic circuit obtained from ${\cal D}_{L,n}$  by a suitable
specialization of the parameters $z_1,\ldots, z_r$ into complex
values. Without loss of generality, we may assume that the number
of internal nodes of ${\cal D}_{L,n}$ is of order $O((L+n)^2)$.
Translating the structure of the directed acyclic graph ${\cal
D}_{L,n}$ into first-order logic one infers easily a formula
$\Psi_{L,n}(S,P,z_1,\ldots,z_r,x_1,\ldots,\ab x_n,\ab y)$ in the
free variables $z_1,\ldots,z_r,x_1,\ab \ldots,\ab x_n,\ab y$ of
the query language $\mathrm{FO}(+,\times,0,1,=,S,P)$ such that
$\Psi_{L,n}(S,P)$ satisfies the following conditions.

\begin{enumerate}
\item[(i)] $\Psi_{L,n}(S,P)$ is prenex, purely existential and of
length $O((L+n)^2)$;

\item[(ii)] interpreting the predicates $S$ and $P$ in
$\Psi_{L,n}$ as the graphs of the addition and the multiplication
of complex numbers, and specializing the variables
$z_1,\ldots,z_r$ into the complex numbers
$\zeta_1,\ldots,\zeta_r$, the formula $\Psi_{L,n}(S,P,\ab
\zeta_1,\ab \ldots,\ab \zeta_r,\ab x_1,\ab \ldots,\ab x_n,\ab y)$
describes the graph of the polynomial of $\C[x_1,\ldots,x_n]$
computed by the arithmetic circuit, obtained from ${\cal D}_{L,n}$
by specializing the scalars $z_1,\ldots, z_r$ into
$\zeta_1,\ldots,\zeta_r$.
\end{enumerate}

Let $m:= 4(L+n)^2+2$. From~\cite{cghmp}, Corollary 3 (see
also~\cite[Lemma 4]{gihe}) we deduce that there exists an
identification sequence $\gamma_1,\ldots, \gamma_m\in \Q^n$ for
the object class $\overline{W_{L,n}}$. Let $\Delta_{L,n}(S,P)$ be
a closed $\mathrm{FO}(+,\times,0,1,=,S,P)$-formula saying that $S$
and $P$ are the graphs of two binary operations which map $\C^2$
into $\C$, that $\gamma_1,\ldots ,\gamma_n$ is an identification
sequence for the object class of applications from $\C^n$ to $\C$,
defined by the $\mathrm{FO}(+,\times,0,1,=,S,P)$-formula
$\Psi_{L,n}(S,P)$ and that this object class is not empty. Without
loss of generality, we may assume that $\Delta_{L,n}{(S,P)}$ has
length $O((L+n)^2)$ and is prenex with a fixed number of
quantifier alternations (which is independent of $L$ and $n$).

We consider now the $\mathrm{FO}(+,\times,0,1,=,S,P)$-formulas
$\Phi_{L,n}(S,\ab P,\ab u_1,\ab \ldots,\ab u_r,\ab x_1,\ab
\ldots,\ab x_n,\ab y)$
 defined by

 $$\displaylines{
 (\exists z_1)\cdots (\exists z_r)(\Psi_{L,n}(S,P, z_1,\ldots,z_r,x_1,\ldots,x_n,y)\land
\cr \hfill{}
 \bigwedge_{1\leq k\leq m}\Psi_{L,n}(S,P, z_1,\ldots,z_r,\gamma_k,u_k))\land \Delta_{L,n}(S,P)\qquad\cr}$$

and $\Omega_{L,n}(S,P,u_1,\ldots,u_r)$
 defined by $$(\exists z_1)\cdots (\exists z_r)
 (\bigwedge_{1\leq k\leq m}\Psi_{L,n}(S,P, z_1,\ldots,z_r,\gamma_k,u_k))
 \land \Delta_{L,n}(S,P).$$

Without loss of generality, we may assume that $\Phi_{L,n}{(S,P)}$
and $\Omega_{L,n}{(S,P)}$ are prenex formulas of length
$O((L+n)^2)$ having a \emph{fixed} number of quantifier
alternations and containing the free variables $u_1,\ldots ,
u_m;x_1,\ldots ,x_n,y$ and $u_1,\ldots ,u_m$, respectively. Let
$\pi:\C^{m+n+1}\rightarrow \C^m$ be the canonical projection which
maps each point of $\C^{m+n+1}$ on its first $m$ coordinates and
let $D$ be a constraint database over the schema $(S,P)$ over the
complex numbers. Suppose that $D$ satisfies the formula
$\Delta_{L,n}(S,P)$. With respect to the database $D$, the formula
$\Phi_{L,n}(S,P,u_1,\ldots,u_r,x_1,\ldots,x_n,y)$ defines a
non-empty constructible subset $A_{L,n}(D)$ of $\C^{m+n+1}$ and
the formula $\Omega_{L,n}(S,P,u_1,\ldots,u_r)$ defines the set
$\pi(A_{L,n}(D))$. Moreover, for any $u\in \pi(A_{L,n}(D))$, the
formula $\Phi_{L,n}(S,\ab P,\ab u,\ab x_1,\ab \ldots ,\ab x_n,\ab
y)$ describes the graph of a $n$-variate polynomial map.
Therefore, it makes sense to consider, for any natural numbers $n$
and $L$, the generalized sample point query associated to the
formula $\Phi_{L,n}(S,\ab P,\ab u_1,\ab \ldots,u_n,\ab x_1,\ab
\ldots ,\ab x_n,\ab y)$. Suppose now that there is given a
\emph{branching-parsimonious} procedure $\cal P$ which evaluates
this family of extended sample point queries. We are now going to
analyze the complexity behaviour of $\cal P$ for this query on the
particular input database $D$, where $S$ and $P$ are interpreted
as the graphs of the sum and the product of complex numbers.

We are now able to state and to prove the main complexity result
of this paper.

\begin{theorem} \label{lowerbound}
Let notations and assumptions be as before. Then the
branching-parsimonious procedure $\mathcal P$ requires sequential
time $2^{\Omega(n)}$ in order to evaluate on input the database
$D$ the extended sample point query associated to the size
$O(n^2)$ first-order formula $\Phi_{n,n}(S,P)$. In particular,
extended sample point queries associated to first-order formulas
with a fixed number of quantifier alternations cannot be evaluated
by branching-parsimonious procedures in polynomial time.
\end{theorem}

\noindent {\bf Proof}. The arguments we are now going to use
follow the general lines of the proofs of \cite[Theorem 5]{gihe}
and \cite[Theorem 4]{cghmp}.

For the moment let us fix the integer parameters $L$ and $n$.

Observe that the closed formula $\Delta _{L,n}(S,P)$ is valid on
the database $D$. Therefore the constructible set
$A_{L,n}:=A_{L,n}(D)$ is nonempty.

Let $B_{L,n}$ and $V_{L,n}$ be the Zariski-closures of $A_{L,n}$
and $\pi(A_{L,n})$ in $\C^{m+n+1}$ and $\C^m$, respectively.

Let $\lambda_{L,n}:= \overline{W}_{L,n} \times \C^n \rightarrow
\C^{m+n+1}$ and $\mu_{L,n}:= \overline{W}_{L,n} \rightarrow
\C^{m}$ be the morphisms of $\Q$-definable affine varieties
defined for $f\in \overline{W}_{L,n}$ and $x\in \C^n$ by
$\lambda_{L,n}(f,x):=(f(\gamma_1),\ldots ,f(\gamma_m),x,f(x))$ and
$\mu_{L,n}(f):=(f(\gamma_1),\ldots ,f(\gamma_m))$. From the
syntactic form of $\Phi_{L,n}(S,P)$ and $\Omega_{L,n}(S,P)$ one
infers immediately that
\[
\lambda_{L,n}(W_{L,n}\times \C^n)=A_{L,n}
\]
and
\[
\mu_{L,n}(W_{L,n})=\pi(A_{L,n})
\]
holds.

Therefore $B_{L,n}$ and $V_{L,n}$ are $\Q$-definable absolutely
irreducible affine varieties and we may consider $\lambda_{L,n}$
and $\mu_{L,n}$ as dominant morphism mapping $\overline{W}_{L,n}
\times \C^n$ into $B_{L,n}$ and $\overline{W}_{L,n}$ into
$V_{L,n}$. Observe that $\overline{W}_{L,n}$ and $V_{L,n}$ form
closed cones in their respective ambient spaces. Since
$\gamma_1,\ldots ,\gamma_m$ were chosen as an identification
sequence for the object class $\overline{W}_{L,n}$, we may
conclude that $\mu_{L,n}:\overline{W}_{L,n} \rightarrow V_{L,n}$
is an injective dominant morphism of closed affine cones, which is
homogeneous of degree one.

Therefore $\mu_{L,n}$ is a finite, bijective and birational
morphism of affine varieties (see, e.g., \cite[I.5.3 Theorem 8 and
proof of Theorem 7]{shafarevich}, \cite[Lemma 4]{gihe} or
\cite[Lemma 5]{cghmp}).

Let $\widetilde{\pi}:\C^{m+n+1} \rightarrow \C^{m+n}$ be the
canonical projection which maps each point of $\C^{m+n+1}$ on its
first $m+n$ coordinates. Then $\widetilde{\pi}\circ \lambda_{L,n}:
\overline{W}_{L,n}\times \C^n \rightarrow V_{L,n}\times \C^n$ is a
finite bijective and birational morphism of affine varieties and
therefore $\lambda_{L,n}:\overline{W}_{L,n}\times \C^n \rightarrow
B_{L,n}$ has the same property. This implies
$\pi(B_{L,n})=V_{L,n}$ and that $B_{L,n}$ represents a rationally
parameterized family of polynomial functions which extends the
family represented by $A_{L,n}$.

Let $K_{L,n}$ the function field over $\C$ of the absolutely
irreducible variety $V_{L,n}$ and let $R_{L,n}$ be the
$\C$-algebra of all rational functions $\theta$ of $K_{L,n}$ such
that for any point $u\in V_{L,n}$ and any place $\varphi :K_{L,n}
\rightarrow \C \cup \{ \infty \}$ whose valuation ring extends the
local ring of the affine variety $V_{L,n}$ at the point $u$, the
value $\varphi(\theta)$ is finite and uniquely determined by $u$.
Thus, for $u\in V_{L,n}$ and $\varphi:K_{L,n} \rightarrow \C \cup
\{ \infty \}$ as above, we may associate to any polynomial
$f:=\sum a_{i_1\ldots i_n}x_1^{i_1}\ldots x_n^{i_n}\in
R_{L,n}[x_1,\ldots ,x_n]$ the polynomial $f(u,x_1,\ldots
x_n):=\sum a_{i_1\ldots i_n}(u)x_1^{i_1}\ldots x_n^{i_n}:=\sum
\varphi(a_{i_1\ldots i_n})x_1^{i_1}\ldots x_n^{i_n}$, which
belongs to $\C[x_1,\ldots ,x_n]$.

Since $\mu_{L,n}:\overline{W}_{L,n} \rightarrow V_{L,n}$ is a
finite, bijective and birational morphism of affine varieties, we
conclude that $R_{L,n}$ contains the coordinate ring of the affine
variety $\overline{W}_{L,n}$ (see e.g. \cite{lang}). This implies
that there exists a polynomial $f_{L,n}\in R_{L,n}[x_1,\ldots
,x_n]\subset K[x_1,\ldots x_n]$ with the following property: for
any $u\in V_{L,n}$, $x\in \C^n$ and $y\in \C$, the point $(u,x,y)$
belongs to $B_{L,n}$ if and only if $f_{L,n}(u,x)=y$ holds.

A branching-free output representation of the extended sample
point query associated to the constructible set $A_{L,n}$ can now
easily be realized by any arithmetic circuit which first computes
all monomial terms of the polynomial $f_{L,n}$ and finally sums
them up.

Therefore, the given branching-parsimonious query evaluation
procedure $\cal P$ produces on input consisting of the database
$D$ and the formula $\Phi_{L,n}(S,P)$ a branching-free
representation of the  extended sample point query associated to
$A_{L,n}$. This branching-free representation is realized by an
essentially division-free single-output arithmetic circuit
$\beta_{L,n}$ with inputs $x_1,\ldots ,x_n$ and scalars
$\theta_1^{(L,n)},\ldots, \theta_{s_{L,n}}^{(L,n)}$ belonging to
$R_{L,n}$ such that $\beta_{L,n}$ computes at its output the
polynomial $f_{L,n}\in R_{L,n}[x_1,\ldots ,x_n]$.

Let $t$ and $\ell_1,\ldots ,\ell_n$ be new variables and let us
now consider the polynomial $g_n:=t \prod_{1\le i\le n } (\ell_i +
x_i)$ defining the constructible object class
\[
\Gamma_n:=\{ \tau \prod_{1\le i\le n } (\lambda_i + x_i)\ \mid \
\tau, \lambda_1,\ldots ,\lambda_n\in \C \}
\]
of $n$-variate complex polynomial functions.

Observe that each element of $\Gamma_n$ has nonscalar sequential
time complexity at most $n$.

Therefore the Zariski-closure $\overline{\Gamma}_n$ of the object
class $\Gamma_n$ is contained in $\overline{W}_{n,n}$.

Observe that $\overline{\Gamma}_n$ is an absolutely irreducible,
$\Q$-definable affine variety. Since
$\mu_{n,n}:\overline{W}_{n,n}\rightarrow V_{n,n}$ is a finite
morphism of irreducible affine varieties, we conclude that
$C_n:=\mu_{n,n}(\overline{\Gamma}_n)$ is an absolutely
irreducible, $\Q$-definable closed affine subvariety of $V_{n,n}$.

Let $E_n$ and $L_n$ be the coordinate ring and the rational
function field over $\C$ of the absolutely irreducible affine
variety $C_n$.

Observe that we may identify $E_n$ with $\C [g_n(t,\ab \ell_1,\ab
\ldots ,\ell_n,\ab \gamma_1),\ab \ldots ,\ab g_n(t,\ab \ell_1,\ab
\ldots ,\ab \ell_n,\ab \gamma_m)]$ and $L_n$ with $\C (g_n(t,\ab
\ell_1,\ab \ldots ,\ab \ell_n,\ab \gamma_1),\ab \ldots ,g_n(t,\ab
\ell_1,\ab \ldots ,\ab \ell_n,\ab \gamma_m))$. Therefore we may
consider $E_n$ as a $\C$-subdomain of the polynomial ring $\C
[t,\ell_1,\ldots ,\ell_n]$ and $L_n$ as a $\C$-subfield of $\C
(t,\ell_1,\ldots ,\ell_n)$.

The rational functions $\theta_1^{(n,n)},\ldots
,\theta_{s_{n,n}}^{(n,n)}$ of the affine variety $V_{n,n}$ may be
not defined on the subvariety $C_n$. Nevertheless, since they
belong to the $\C$-algebra $R_{n,n}$, one verifies easily that
there exist rational functions $\sigma_1^{(n)},\ldots
,\sigma_{s_{n,n}}^{(n)}$ of the affine variety $C_n$ satisfying
the following condition: for any point $u\in C_n$ and any place
$\psi: L_n \rightarrow \C \cup \{ \infty \}$ whose evaluation ring
extends the local ring of $C_n$ at the point $u$, the values of
$\psi$ at $\sigma_1^{(n)},\ldots ,\sigma_{s_{n,n}}^{(n)}$ are
given by $\psi(\sigma_1^{(n)})=\theta_1^{(n,n)}(u),\ldots
,\psi(\sigma_{s_{n,n}}^{(n)})=\theta_{s_{n,n}}^{(n,n)}(u)$ and
therefore finite and uniquely determined by $u$.

In particular, the rational functions $\sigma_1^{(n)},\ldots
,\sigma_{s_{n,n}}^{(n)}\in L_n$ are integral over the $\C$-algebra
$E_n$ and hence contained in the polynomial ring
$\C[t,\ell_1,\ldots ,\ell_n]$ (see, e.g., \cite{lang}). Therefore
we may consider $\sigma_1^{(n)},\ldots ,\sigma_{s_{n,n}}^{(n)}$ as
polynomials in the variables $t$ and $\ell_1,\ldots ,\ell_n$ (i.e.
as elements of $\C[t,\ell_1,\ldots ,\ell_n]$).

>From $g_n=t\prod_{1\le i\le n} (\ell_i + x_i)$, we infer the
identities
\[
g_n(0,\ell_1,\ldots ,\ell_n,\gamma_1)=0,\ldots ,
g_n(0,\ell_1,\ldots ,\ell_n,\gamma_m)=0.
\]

Since the polynomials $\sigma_1^{(n)},\ldots
,\sigma_{s_{n,n}}^{(n)}$ depend integrally from $g_n(t,\ab
\ell_1,\ab \ldots ,\ell_n,\ab \gamma_m),\ldots ,\ab g_n(t,\ab
\ell_1,\ab \ldots ,\ab \ell_n,\ab \gamma_m)$, we deduce now easily
that the polynomials $\sigma_1^{(n)}(0,\ab \ell_1,\ab \ldots ,\ab
\ell_n),\ab \ldots ,\ab \sigma_{s_{n,n}}^{(n)}(0,\ab \ell_1,\ab
\ldots ,\ab \ell_n)$ do not depend on the variables $\ell_1,\ab
\ldots ,\ab \ell_n$, i.e., they belong to $\C$.

Let now $\widetilde{\beta}_n$ be the division-free arithmetic
circuit with scalars in $\C[t,\ab \ell_1,\ab \ldots ,\ab \ell_n]$
obtained by replacing in the circuit $\beta_{n,n}$ the scalars
$\theta_1^{(n,n)},\ab \ldots ,\ab \theta_{s_{n,n}}^{(n,n)}$ by the
polynomials $\sigma_1^{(n)},\ldots ,\sigma_{s_{n,n}}^{(n)}$.


One verifies easily that the circuit $\widetilde{\beta}_n$
computes the polynomial
\[
g_n=t\prod_{1\le i\le n} (\ell_i + x_i)=\sum _{\delta_1,\ldots
,\delta_n,\varepsilon_1,\ldots , \varepsilon_n \in \{ 0,1\}\atop{
{\delta_1+\varepsilon_1=1,\ldots,\delta_n + \varepsilon_n=1 }}} t
\ell_1^{\delta_1}\ldots \ell_n^{\delta_n}x_1^{\varepsilon_1}\ldots
x_n^{\varepsilon_n}.
\]

Let $v_1,\ldots ,v_{s_{n,n}}$ be new variables. From the directed
acyclic graph structure of $\widetilde{\beta}_n$ (or
$\beta_{n,n}$) one deduces immediately that for each
$(\delta_1,\ldots ,\delta_n)\in \{0,1\}^n$ there exists a
polynomial $Q_{(\delta_1,\ldots ,\delta_n)}^{(n)}\in \Q[v_1,\ldots
,v_{s_{n,n}}]$ satisfying the condition $Q_{(\delta_1,\ldots
,\delta_n)}^{(n)}(\sigma_1^{(n)},\ldots
,\sigma_{s_{n,n}}^{(n)})=t\ell_1^{\delta_1}\ldots
\ell_n^{\delta_n}$. Let $Q_n:\C ^{s_{n,n}} \rightarrow \C^{2^n}$
the polynomial map defined by $Q_n:=(Q_{(\delta_1,\ldots
,\delta_n)}^{(n)};(\delta_1,\ldots ,\delta_n)\in \{0,1\}^n)$.

Consider now an arbitrary integer $1\le \rho \le 2^n$ and let
$\lambda_{\rho, 1}:=\rho ^{2^0},\ldots ,\lambda_{\rho, n}:=\rho
^{2^{n-1}}$, $\lambda_\rho := (\lambda_{\rho, 1},\ldots
,\lambda_{\rho, n})$ and $\alpha_\rho ^{(n)}:\C \rightarrow
\C^{s_{n,n}}$ and $\beta_\rho ^{(n)}:\C \rightarrow \C^{2^n}$ be
the parameterized algebraic curves defined for $\tau\in \C$ by
\[
\alpha_\rho ^{(n)}(\tau):=(\sigma_1^{(n)}(\tau,
\lambda_\rho),\ldots ,\sigma_{s_{n,n}}^{(n)}(\tau, \lambda_\rho))
\]
and
\[
\beta_\rho ^{(n)}(\tau):=(\tau \lambda_{\rho_,1}^{\delta_1}\ldots
\lambda_{\rho_,n}^{\delta_n};\ (\delta_1,\ldots ,\delta_n)\in
\{0,1\}^n)=(\tau\rho^j;\ 0\le j<2^n).
\]

Observe that the functional identity
\begin{equation} \label{functional}
\beta_\rho^{(n)}=Q_n\circ \alpha_\rho^{(n)}
\end{equation}
is valid.

Since the polynomials $\sigma_1^{(n)}(0,\ell_1,\ldots
,\ell_n),\ldots ,\sigma_{s_{n,n}}^{(n)}(0,\ell_1,\ldots ,\ell_n)$
do not depend on the variables $\ell_1,\ldots ,\ell_n$, there
exists a point $a_n\in \C^{s_{n,n}}$, independent on $\rho$, such
that $\alpha_\rho ^{(n)}(0)=a_n$ holds.

Let us denote the derivatives of $\alpha_\rho^{(n)}$,
$\beta_\rho^{(n)}$ and $Q_n$ by
$\frac{d\alpha_\rho^{(n)}}{d\tau}$,
$\frac{d\beta_\rho^{(n)}}{d\tau}$ and $DQ_n$. Furthermore let
$\omega_\rho^{(n)}:=\frac{d\alpha_\rho^{(n)}}{d\tau}(0)\in
\C^{s_{n,n}}$, let $\eta_n:\C ^{s_{n,n}} \rightarrow \C^{2^n}$ be
the $\C$-linear map defined by $\eta_n:=(DQ_n)(a_n)$ and observe
that $\frac{d\beta_\rho^{(n)}}{d\tau}(0)=(\rho^j;\ 0\le j<2^n)$
holds. Applying the chain rule to (\ref{functional}), we infer the
following identities:


\[
\begin{tabular}{lll}
$(\rho^j;\ 0\le j<2^n)$&$=$&$\frac{d\beta_\rho^{(n)}}{d\tau}(0)=
(DQ_n)(\alpha_\rho^{(n)}(0)(\frac{d\alpha_\rho^{(n)}}{d\tau}(0))$\\
&$=$ &$(DQ_n)(a_n)(\omega_\rho^{(n)}) =\eta_n(\omega_\rho^{(n)}).$
\end{tabular}
\]

Since $(\rho^j)_{1\le \rho \le 2^n, 0\le j<2^n}$
is a nonsingular Vandermonde matrix, we conclude now that the
image of the $\C$-linear map $\eta_n:\C ^{s_{n,n}} \rightarrow
\C^{2^n}$ contains $2^n$ linear independent points. Therefore
$\eta_n$ is surjective. This implies $s_{n,n}\ge 2^n$.

Therefore the arithmetic circuit $\beta_{n,n}$, which represents
the output produced by the procedure $\mathcal P$ on input
consisting of the database $D$ and the formula $\Phi_{n,n}$,
contains at least $2^n$ scalars.

This implies that the nonscalar size of the circuit $\beta_{n,n}$
is at least $2^{\frac{n}{2}}-n-1$. In conclusion, the procedure
$\mathcal P$ requires $2^{\Omega(n)}$ sequential time in order to
produce the output $\beta_{n,n}$ on input consisting of the data
base $D$ and the size $O(n^2)$ formula $\Phi_{n,n}$. \qed

The main outcome of Theorem \ref{lowerbound} and its proof can be
paraphrased as follows: \emph{constraint database theory applied
to quite natural computation tasks, as, e.g.,
branching-parsimonious interpolation of low complexity
polynomials, leads necessarily to non-polynomial sequential time
lower bounds.}

In view of the $P_\R \not= NP_\R$ conjecture in the algorithmic
model of Blum--Shub--Smale
 over the real and complex numbers, it seems unlikely that this
worst case complexity behavior can be improved substantially if we
drop some or all of our previously introduced requirements on
queries and their output representations. Nevertheless we wish to
stress that these requirements constitute a fundamental technical
ingredient for the argumentation in the proof of Theorem
\ref{lowerbound}.

\section{Conclusion and future research on the complexity of query
evaluation}\label{section5}

\bartq{Still from CDB} In this paper, we have emphasized the
importance of \emph{data structures} and their effect on the
complexity of quantifier elimination.

However, the intrinsic inefficiency of quantifier-elimination
procedures represents a bottle-neck for real-world implementations
of constraint database systems. As we have argued, it is unlikely
that constraint database systems that are based on general purpose
quantifier-elimination algorithms will ever become efficient.
Also, restriction to work with linear data, as in most existing
constraint database systems~\cite[Part IV]{cdbook}, will also not
lead to more efficiency. A promising direction is the study of a
concept like the \emph{system degree}, that has shown to be a
fruitful notion for the complexity analysis of quantifier
elimination in elementary geometry and has been implemented in the
software package (polynomial equation solver "Kronecker", see
\cite{gls}). In the context of query evaluation in constraint
databases, the notion of system degree is still unsatisfactory
since it is determined both by the query formula and the
quantifier-free formulas describing the input database relations.
It is a task for future constraint database research to develop a
well-adapted complexity invariant in the spirit of the system
degree in elimination theory.

Another direction of research is the study of query evaluation for
first-order languages that capture certain genericity classes. For
example, the first-order logic \fobetween\ has point variables
rather than being based on real numbers and it captures the
fragment of first-order logic over the reals that expresses
queries that are invariant under affine transformations of the
ambient space~\cite{jmd}. Although a more efficient complexity of
query evaluation in this language cannot be expected, it is
interesting to know whether languages such as \fobetween\ have
quantifier elimination themselves (after an augmentation with
suitable predicates).

\bibliographystyle{plain}

\clearpage
\section*{Appendix A: Language and tools from algebraic geometry.}

 Let $k$ be the field $\Q$, $\R$ or $\C$.  We assume $k$ to be
``effective" with respect to arithmetic operations as
addition/subtraction, multiplication/division. Let $\overline{k}$
be an algebraically closed field containing $k$ (in the sequel we
shall call such a field {\em an algebraic closure of $k$}).

We denote by $\N$ the set of natural numbers and by $\Zpos$ the
set of nonnegative integers.\bartq{remove?}
\medskip

Fix $n\in\Zpos$ and let $X_0\klk X_n$ be indeterminates over $k$.
We denote by $\A^n:=\A^n(\overline{k})$ the $n$--dimensional
affine space and by $\Pe^n:=\Pe^n(\overline{k})$
\bartq{$\overline{k}\rightarrow\C$?} the $n$--dimensional
projective space over $\overline{k}$. The spaces $\A^n$ and
$\Pe^n$ are thought to be endowed with their respective Zariski
topologies over $k$ and with their respective sheaves of
$k$--rational functions with values in $\overline{k}$. Thus the
points of $\A^n$ are elements $(x_1\klk x_n)$ of $\overline{k}$
and the points of $\Pe^n$ are (non uniquely) represented by
non--zero elements $(x_0\klk x_n)$ of $\overline{k}^{n+1}$ and
denoted by $(x_0:\dots: x_n)$. The indeterminates $X_1\klk X_n$
are considered as the coordinate functions of the affine space
$\A^n$. The coordinate ring (of polynomial functions) of $\A^n$ is
identified with the polynomial ring $k[X_1\klk X_n]$. Similarly we
consider the (graded) polynomial ring $k[X_0\klk X_n]$ as the
projective coordinate ring of $\Pe^n$. Consequently we represent
rational functions of $\Pe^n$ as quotients of homogeneous
polynomials of equal degree belonging to $k[X_0\klk X_n]$. Let
$F_1\klk F_s$ be polynomials which belong to $k[X_1\klk X_n]$ or
are homogeneous and belong to $k[X_0\klk X_n]$. We denote by
$\{F_1=0\klk F_s=0\}$ or $V(F_1\klk F_s)$ the algebraic set of
common zeroes of the polynomials $F_1\klk F_s$ in $\A^n$ and
$\Pe^n$ respectively. We consider the set $V:=\{F_1=0\klk F_s=0\}$
as (Zariski--)closed (affine or projective) subvariety of its
ambient space $\A^n$ or $\Pe^n$ and call $V$ the affine or
projective variety defined by the polynomials $F_1\klk F_s$. We
think the variety $V$ to be equipped with the induced Zariski
topology and its sheaf of rational functions. The irreducible
components of $V$ are defined with respect to its Zariski topology
over $k$ . We call $V$ irreducible if $V$ contains a single
irreducible component and equidimensional if all its irreducible
components have the same dimension. The dimension $dim\,V$ of the
variety $V$ is defined as the maximal dimension of all its
irreducible components. If $V$ is equidimensional we define its
(geometric) degree as the number of points arising when we
intersect $V$ with $dim\,V$ many generic (affine) linear
hyperplanes of its ambient space $\A^n$ or $\Pe^n$. For an
arbitrary closed variety $V$ with irreducible components
$\mathcal{C}_1\klk\mathcal{C}_t$ we define its degree as $\deg
V:=\deg \mathcal{C}_1+\cdots +\deg\mathcal{C}_t$. With this
definition of degree the intersection of two closed subvarieties
$V$ and $W$ of the same ambient space satisfies the B{\'e}zout
inequality
$$\deg V\cap W\le \deg V\deg W$$.

We denote by $k[V]$ the affine or (graded) projective coordinate
ring of the variety $V$. If $V$ is irreducible we denote by $k(V)$
its field of rational functions. In case that $V$ is a closed
subvariety of the affine space $\A^n$ we consider the elements of
$k[V]$ as $\overline{k}$--valued functions mapping $V$ into
$\overline{k}$. The restrictions of the projections $X_1\klk X_n$
to $V$ generate the coordinate ring $k[V]$ over $k$ and are called
the coordinate functions of $V$. The data \bartq{What is
\emph{data} here?} of $n$ coordinate functions of $V$ fixes an
embedding of $V$ into the affine space $\A^n$. Morphisms between
affine and projective varieties are induced by polynomial maps
between their ambient spaces which are supposed to be homogeneous
if the source and target variety is projective.

Replacing the ground field $k$ by its algebraic closure
$\overline{k}$, we may apply all this terminology again. In this
sense we shall speak about the Zariski topologies and coordinate
rings over $\overline{k}$ and sheaves of $\overline{k}$--rational
functions. In this more general context varieties are defined by
polynomials with coefficients in $\overline{k}$. If we want to
stress that a particular variety $V$ is defined by polynomials
with coefficients in the ground field $k$, we shall say that $V$
is $k$--{\em definable} or $k$--{\em constructible}. The same
terminology is applied to any set determined by a (finite) boolean
combination of $k$--definable closed subvarieties of $\A^n$ or
$\Pe^n$. By a {\em constructible} set we mean simply a
$\overline{k}$--constructible one. Constructible and
$k$--constructible sets are always thought to be equipped with
their corresponding Zariski topology. In case of $k:=\Q$ and
$\overline{k}:=\C$ we shall sometimes also consider the {\em
euclidean} (i.e. ``{\em strong}") topology of $\A^n$ and $\Pe^n$
and their constructible subsets.\bartq{Is this necessary?}
\medskip

The rest of our terminology and notation of algebraic geometry and
commutative algebra is standard and can be found in \cite{lang},
\cite{shafarevich}, and in \cite{atiyah}.



\end{document}